\documentclass[review]{elsarticle}

\usepackage{lineno,hyperref}
\usepackage{caption}
\usepackage{subcaption}
\usepackage{tabularx}
\usepackage{array}
\graphicspath{{./pictures/}} 

\usepackage{tikz}
\usetikzlibrary{shapes.geometric,arrows,positioning}

\newcommand{\ORNL}{Materials Science and Technology Division,
 Oak Ridge National Laboratory, Oak Ridge, Tennessee 37831-6138, USA}

\newcommand{\UTK}{Department of Materials Science, University of Tennessee, Knoxville,TN,USA}

\begin{document}

\title{Kinetic Activation-Relaxation Technique and Self-Evolving Atomistic Kinetic Monte Carlo: Comparison of on-the-fly kinetic Monte Carlo algorithms} 


\author{Laurent Karim B\'{e}land}
\author{Yury N. Osetsky}
\author{Roger Stoller}
\address{\ORNL}
\author{Haixuan Xu \corref{mycorrespondingauthor}}
\cortext[mycorrespondingauthor]{Corresponding author}
\ead{xhx@utk.edu}
\address{\UTK}

\date{\today}

\begin{abstract}

We present a comparison of the kinetic Activation-Relaxation Technique (k-ART) and the Self-Evolving Atomistic Kinetic Monte Carlo (SEAKMC), two off-lattice, on-the-fly kinetic Monte Carlo (KMC) techniques that were recently used to solve several materials science problems. We show that if the initial displacements are localized the dimer method and the Activation-Relaxation Technique \emph{nouveau} provide similar performance.  We also show that k-ART and SEAKMC, although based on different approximations, are in agreement with each other, as demonstrated by the examples of 50 vacancies in a 1950-atom Fe box and of interstitial loops in 16000-atom boxes. Generally speaking, k-ART's treatment of geometry and flickers is more flexible, e.g. it can handle amorphous systems, and rigorous than SEAKMC's, while the later's concept of active volumes permits a significant speedup of simulations for the systems under consideration and therefore allows investigations of processes requiring large systems that are not accessible if not localizing calculations.

\end{abstract}


%\pacs{1}

\maketitle

\section{Introduction}
On-the-fly kinetic Monte Carlo (KMC) methods are increasingly popular numerical tools to study diverse structures and processes in materials science. Some examples include defect diffusion in SiC \cite{jiang2014accelerated}, Fe \cite{beland2011kinetic,joly2012optimization,brommer2012comment,xu2013cascade,xu2011simulating,xu2012self,xu2013solving,chill2014molecular,BrommerArxiv}, Ag \cite{kara2009off,nandipati2009parallel,latz2012three}, Pt \cite{jonsson2011simulation}, Al \cite{henkelman2001long}, Cu \cite{trushin2005self,pedersen2009long}, Pd on MgO \cite{xu2008adaptive}, methanol on Cu \cite{xu2009adaptive}, CO on ice \cite{karssemeijer2012long}, binary alloys \cite{bleda2008calculations}, \emph{c}-Si \cite{el2008kinetic,beland2011kinetic,joly2012optimization} and \emph{a}-Si \cite{beland2011kinetic,joly2012optimization,mousseau2012activation,joly2013contribution}, as well as other more complex processes such as surface erosion, thin film growth \cite{scott2011atomistic}, sputtering of Au surfaces \cite{scott2013sputtering} and post-ion bombardement annealing in \emph{c}-Si \cite{beland2013replenish,beland2013long}. This enthusiasm is justified by these methods ability to handle off-lattice positions, executing high-energy barriers with ease at any temperature (thus potentially reaching long simulation times), while requiring little \emph{a priori} knowledge of the nature of the physical processes involved in the kinetics.

Sampling saddle points surrounding any given minimum in the potential energy landscape very efficiently is crucial to the success of such an endeavor. Indeed, since the list of processes governing the kinetics is built on-the-fly, these methods must quickly evaluate all the relevant exit states linked to a given configuration. Some methods, such as the Self-Evolving Atomistic KMC \cite{xu2011simulating,xu2012self} (SEAKMC), sample a relatively small number of such transition states each step (several tens or hundreds), while self-learning methods, such as the kinetic Activation Relaxation Technique \cite{el2008kinetic,beland2011kinetic,mousseau2012activation} (k-ART), sample a larger number of states when encountering a configuration never observed in the past, while sampling a smaller number when encountering a configuration that was previously visited (adding events stored in memory to the list of processes that must be taken into account). In all cases, one can hardly understate the importance of efficiently finding saddle points.

Current methods model the kinetics using harmonic transition state theory \cite{glasstone1941theory,vineyard1957frequency,voter2002extending} (hTST), choosing a constant reaction rate pre-factor. In this context, saddle searches are conducted using either minimum-mode following methods or high temperature molecular dynamics combined with the Nudged Elastic Band \cite{henkelman2000climbing,chill2014molecular} method. The latter requires prior knowledge to be able to detect when the temperature assisted MD \cite{so2000temperature} has crossed a saddle point. The minimum-mode following methods have been more widely used in the on-the-fly KMC context. Thus, this article will focus on these.

Minimum-mode following methods include the Dimer method \cite{henkelman1999dimer}, Raleigh-Ritz minimization \cite{horn2012matrix}, the hybrid eigenvector following method \cite{munro1999defect} and the activation relaxation technique \emph{nouveau} \cite{barkema1996event,malek2000dynamics} (ARTn). While some comparisons between these techniques have been made in the past \cite{olsen2004comparison,heyden2005efficient,machado2011optimized}, prescriptions concerning the optimal choice of method for studying a given system are lacking. Indeed, most problems of interest have energy landscape that can significantly differ from the idealized potential energy landscapes of the simple defects that are used when designing these numerical methods. 

In addition to the choice of saddle-search method, the on-the-fly kMC techniques use different sets of approximations in order to accelerate the simulations. While molecular dynamics (MD) \cite{xu2013cascade,xu2013solving}, and experiments \cite{beland2013replenish} indicate that these simulations captured the relevant physics, there are few direct comparison of these accelerated methods, which could assess the validity of these approximations. Such a comparison has been attempted in a model-system: the aggregation of 50 vacancies in bcc-Fe. This system was simulated using Autonomous Basin Climbing (ABC) \cite{fan2011mechanism}, k-ART \cite{brommer2012comment,BrommerArxiv}, SEAKMC \cite{xu2013cascade} and Molecular-Dynamics Saddle-Search Adaptive KMC (MDSS-aKMC) \cite{chill2014molecular}. At 323K, all 50 vacancies in a 2000-atom box were found by SEAKMC to aggregate into clusters within 0.01 ms. k-ART estimated this time to be 0.1ms, while ABC estimated it at about 1 day. MDSS-aKMC simulated this process at 423K and found an aggregation time of roughly 1 ms. It was suggested that ABC's large estimate was related to the use of a single barrier, which under-samples the system's transition states\cite{brommer2012comment}. While the k-ART and SEAKMC simulations where performed using very similar parameters, their respective predictions vary by one order of magnitude. To our knowledge, while there were attempts to establish the accuracy of this class of methods \cite{bhute2013building,bhute2013accuracy,chill2014molecular}, no other comparison of such accelerated methods exists.

In this article, we first compare the two most widely methods used in modern on-the-fly kMC, the dimer method and ARTn. We concentrate on four case studies: the Fe mono-vacancy, the Fe self-intersitial atom, the Fe 50-vacancy system and FeCr dislocation loops. Secondly, we perform a systematic comparison of k-ART and SEAKMC for the 50-vacancy problem and for the transformation of two $1/2<111>$ interstitial-loops in Fe into single $<100>$ or $1/2<111>$ loops.

% Define block styles
\tikzstyle{decision} = [diamond, draw, fill=blue!20, 
    text width=8em, text badly centered, node distance=3cm, inner sep=0pt]
\tikzstyle{block} = [rectangle, draw, fill=blue!20, 
    text width=8em, text centered, rounded corners, minimum height=4em]
\tikzstyle{line} = [draw, -latex']
\tikzstyle{cloud} = [draw, ellipse,fill=red!20, node distance=3cm,
    minimum height=2em]

\begin{figure}
\centering
\begin{tikzpicture}[node distance = 2cm, auto]

    % Place nodes
    \node [block] (init) {Global minimization};
    \node [block, below of=init] (identify) {Identification of local configurations};
    \node [block, below of=identify] (evaluate) {Association of events to local configurations};
    \node [decision, below of=evaluate] (decide) {Execution of one event at random; advance clock};
    % Draw edges
    \path [line] (init) -- (identify);
    \path [line] (identify) -- (evaluate);
    \path [line] (evaluate) -- (decide);
    \path [line] (decide.east) |- ([yshift=+1cm] init.north) -| (init.north);
\end{tikzpicture}
\caption{Simplified workflow for on-the-fly kMC algorithms, such as k-ART and SEAKMC.}
\label{Fig:simple_flowchart}
\end{figure}
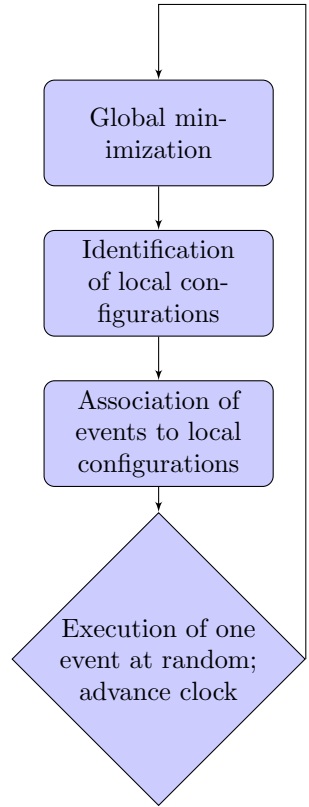

% \usepackage{graphicx}
\begin{table}[t]
\caption{The approximations that k-ART and SEAKMC use to accelerate on-the-fly off-lattice kMC. They correspond to the steps described by the flowchart of Fig.~\ref{Fig:simple_flowchart}. There are no major approximations to the global minimization.}
\resizebox{0.5\textwidth}{!}{%
\begin{tabular}{| m{0.23\textwidth} m{0.23\textwidth} | }
\hline
\multicolumn{2}{|c|}{Identification of local configurations} \\ \hline
\multicolumn{1}{|c}{k-ART} & \multicolumn{1}{c|}{SEAKMC} \\ \hline
Local configurations are well described by topology of connectivity graph. & Local configs are well described by distances of atoms from the perfect lattice. \\ \hline
\end{tabular}
}
\resizebox{0.5\textwidth}{!}{%
\begin{tabular}{| m{0.23\textwidth} m{0.23\textwidth} | }
\hline
\multicolumn{2}{|c|}{Association of events to local configurations} \\ \hline
\multicolumn{1}{|c}{k-ART} & \multicolumn{1}{c|}{SEAKMC} \\ \hline
Events can be catalogued using topology. & Only a subset of atoms are necessary to handle elastic effects on saddle- point. \\
 & \\
Events can be mapped using graph canonical representation. & Sometimes neglect checks for duplicity of events. \\
 & \\
Most important events are sampled when encountering a new topology. & Neglect detailed balance. (Formally, a large number of searches will still lead to micro-reversibility.) \\
 & \\
Expansive, but incomplete, catalog is rarely resampled. & Catalog is resampled with uncorrelated random searches at every KMC step to compensate limited number of searches. \\
 & \\
hTST & hTST \\
 & \\
Constant prefactor & Constant prefactor \\ \hline
\end{tabular}

}

\resizebox{0.5\textwidth}{!}{%
\begin{tabular}{| m{0.23\textwidth} m{0.23\textwidth} | }
\hline
\multicolumn{2}{|c|}{Event execution} \\ \hline
\multicolumn{1}{|c}{k-ART} & \multicolumn{1}{c|}{SEAKMC} \\ \hline
Poisson process. & Poisson process. \\
 & \\
Uncorrelated processes. & Uncorrelated processes. \\
 & \\
Mean-passage time correctly describes kinetics of flickers &  \\ \hline
\end{tabular}

}

\label{Tab:Approximations}
\end{table}

\section{Methods}

\subsection{Similarities and Differences of k-ART and SEAKMC}
k-ART and SEAKMC follow a similar workflow, which is illustrated in Fig.~\ref{Fig:simple_flowchart}. The four steps described in the chart are the basis of kMC: global minimization, identification of local configurations, association of events to local configurations and the execution of one event at random (advancing the clock). For each of these steps, the two methods use a different set of approximations to accelerate simulations. These are summarized in table \ref{Tab:Approximations}. More detailed descriptions of these algorithms can be found in numerous references \cite{el2008kinetic,beland2011kinetic,xu2011simulating,mousseau2012activation,joly2012optimization,xu2012self}. The global minimization is done using FIRE \cite{bitzek2006structural} in k-ART and conjugate gradient in SEAKMC. This step is not a bottleneck for the system sizes that we are studying (2000 to 20000 atoms). 

k-ART and SEAKMC identify local configurations differently. k-ART uses a automorphism identification software package, NAUTY \cite{mckay1981practical,mckay2007nauty}, to identify connectivity graphs formed by an atom's neighboring atoms\cite{el2008kinetic}. This procedure is lightweight, only requiring a small fraction of the overall computational effort. Event generation will be concentrated on atomic neighborhoods which topology does not correspond to the crystal. One should note that while a cut-off is used to determine topology, all saddle-searches are free to displace all atoms in the system. Furthermore, event search for previously visited topologies will not be attempted from scratch, but will rather recycle and refine events that were found in the past. Meanwhile, SEAKMC simply identifies defects by comparing the atomic positions of atoms to those of a perfect lattice or can be based on energy or stress criteria. The topology-based approach is more flexible and demands less \emph{a priori} knowledge of the system. On the other hand, by precisely identifying defects, one can make saddle-searches more effective by centering them on these positions, which can increase the success rate of finding interesting events.

k-ART and SEAKMC also differ when associating events to local configurations. The former takes full advantage of the topology analysis to store and catalog events for latter use. Indeed, events are generated from scratch when new topologies are discovered, and are recycled and refined when these topologies are revisited. Generation of completely new events for known topologies is also done periodically, going as $log_{10}(n)$, where $n$ is the number of times a topology is encountered during a KMC run. When recycling and refining, these events associated with topologies are mapped to specific atomic configurations through geometric transformations that make use of the canonical representation of graphs generated by NAUTY. While this procedure works well for many classes of events, the canonical representation of atomic neighborhoods may leave ambiguities concerning the reconstruction of events, resulting in bad reconstruction of events. In these cases, configurations are still refined to the nearest saddle-point, which may not be related to the original event, but that still has physical meaning. The correspondance of topology to geometry is one of the main approximations of k-ART. Also, in order for k-ART to produce good kinetics, most of the important transition states must be sampled when first encountering a topology, since it will not be resampled often in the future. This is particularly important when using the basin auto-constructing Mean-Rate Method (bacMRM) \cite{puchala2010energy,beland2011kinetic} to handle flickers, as the states in the basin will not be revisited.

In the case of SEAKMC, the main approximation is the concept of active volumes. The algorithm identifies a region surrounding each defect where event searches are constrained. This significantly reduces the computational cost of force evaluations and generally makes saddle-search and transition state sampling more efficient \cite{xu2011simulating,pedersen2011efficient,xu2012self,pedersen2014bowl}, at the risk of neglecting some very long-range interactions, since the radius of active volumes are at least a few lattice constants. Furthermore, investigation of single vacancy diffusion in bcc Iron\cite{xu2013cascade} shows that adding duplicate saddle points to the catalog does not change the defect dynamics or randomness of the diffusion. Therefore, SEAKMC sometime does not verify that duplicate events are associated with the same defect, which can lead to inaccuracies in the total rate calculation. However, while errors in multiplicity affect rates linearly, this effect is usually small compared to errors on barrier height, that effect rates exponentially. Also, SEAKMC does not garantee detailed balance, i.e. it does not assure that the events necessary to reverse a trajectory are included in the catalog. While this formally excludes proper simulations at thermodynamic equilibrium, kinetics of off-equilibrium systems should be correctly captured. In principle, with more saddle point sampling, this issue could be resolved. However, for complex defects, obtaining the required sampling is challenging. Finally, SEAKMC will deliberately sample an incomplete catalog of events for each defect at each KMC step, correcting for this by flushing this catalog when an event is executed and resampling a new uncorrelated catalog.

In addition, both of these methods are based on harmonic Transition State Theory (hTST) and constant pre-factors for transition rates. However, these two approximations are not inherent to these off-lattice KMCs. In principle, one could calculate pre-factors using harmonic or quasi-harmonic theories. One could even perform free energy calculations of these barriers, albeit at a great computational cost. When executing events, both methods also treat transitions as uncorrelated Poisson processes, the basic approximation of kMC. Moreover, k-ART approximates super-basins of states connected by small barriers (flickers) by postulating that the mean-passage time adequately captures the relevant physics. This approximation is implemented through the bacMRM.

\subsection{Set-up of Saddle-search Methods Comparison} \label{Sub:method-saddle}

In order to compare the dimer method and ARTn, we implement both in the Self-evolving atomistic kinetic Monte Carlo (SEAKMC). 
One of SEAKMC's main features is the use of the active volume concept, which are local regions encompassing the defects in the system of interest that limit the region where atoms will be displaced during the saddle search. This concept is very useful in mostly crystalline systems where defects can be readily identified. By limiting the region where force calculations need to be performed, SEAKMC can execute saddle searches much faster than in the case where force calculations are performed across the whole system. By ensuring that both the dimer method and ARTn perform searches constrained to the same subset of atoms and the same implementation of the potentials, we try to make this comparison as fair as possible. Also, we adjusted the Dimer and ARTn search parameters to make them as efficient as possible for each system.

We use a recent implementation of ARTn, which uses improvements described in several studies \cite{marinica2011energy,cances2009some,machado2011optimized,mousseau2012activation}. However, we do not use the Direct Inversion in the Iterative Subspace \cite{csaszar1984geometry}, but instead use the Newton-like method described in Ref. \cite{cances2009some} to control the step-size when converging towards the saddle-point.

We also explore the effect of the initial deformation of the system on the overall performance of the saddle-search. This is known to affect the convergence and sampling efficiency of this class of methods \cite{pedersen2011efficient,pedersen2014bowl}. In the case of the dimer method, we try randomly deforming the system both locally, constraining the initial deformation within 2 lattice-parameters of the point defects, and globally, deforming the entire Active Volume. In ARTn, we controlled the maximum number of steps where the system is pushed in a direction determined at random at the start of the search. If the maximum number of steps is reached or if the softest eigenmode becomes imaginary (negative eigenvalue), then ARTn starts following this eigendirection.

We use the Ackland (2004)\cite{ackland2004development} semi-empirical potential for Fe. In the case of FeCr, we use a two-band potential where Fe interactions are based on Ackland (2004), as developed by Olsson \cite{olsson2005two} . Saddles are converged within a precision of 0.0285 eV/\AA~ (0.01 eV per lattice parameter). All benchmarks use an OpenMP parallelization over 12 cores on 2x X5650 Intel Xeon CPUs  (at 2.67 GHz).

First, we compare the performance of both frameworks in the case of a mono-vacancy in a 10-lattice-parameter Fe box (1999 atoms). Diffusion in such a system is characterized by a two-step process, which includes a stable state and a metastable state \cite{malerba2010comparison}. We launch 120 saddle searches from both states and a use a 2.7 lattice-parameter radius spherical active volume. In order to better understand the saddle-search process, we also ran modified versions of the algorithms on this system. For instance, we ran a version of FIRE where a Lanczos evaluation of the softest mode is performed at every step and where the force is inverted in that eigendirection. We also ran a variant of the dimer method that minimizes $\mathbf{F}=\mathbf{F}_T-2\mathbf{F}_p$ both when the curvature is concave and convex, where $\mathbf{F}_T$ is the total force on the dimer and $\mathbf{F}_p$ is the force parallel to the dimer. In the standard dimer algorithm, this quantity is minimized only when the curvature is convex. 

Second, we compare the performance of both frameworks in the case of a (110) dumbbell SIA in a 10-lattice-parameter Fe box (2001 atoms). We launch 120 saddle searches and use a 4.0 lattice-parameter radius spherical active volume.

Third, we compare the performance of both frameworks in the case of 50 vacancies in a 10-lattice-parameter box (1950 atoms). We launch 120 saddle searches from one of the states generated by a SEAKMC run, using a 2.7 lattice-parameter radius spherical active volume.

Fourth, we compare the performance of both frameworks in the case of two 37-atom interstitial loops in a 20-lattice-parameter Fe box (16074 atoms). We launch 120 saddle searches from one of the states generated by a SEAKMC run, using a 4.5 lattice-parameter radius spherical active volume.

Fifth, we compare the performance of both frameworks in the case of two 37-atoms interstitial loops in a 20-lattice-parameter FeCr (10\% Cr) box (16074 atoms). We launch 120 saddle searches from one of the states generated by a SEAKMC run. We use a 4.5 lattice-parameter radius spherical active volume.

\subsection{Set-Up of on-the-fly KMC comparison}
\subsubsection{50-vacancy Problem} 

We use the same 50-vacancy system as for the saddle-search comparison. We simulate the kinetics of aggregation at 323K, with a constant pre-factor of $5 \times 10^{12} s^{-1}$. We run simulations until the potential energy of the system reaches -7760 eV, which roughly corresponds to the energy when half the mono-vacancies have aggregated.

Previous comparisons of k-ART and SEAKMC looked at simulations of the aggregation of 50 Fe vacancies \cite{xu2013cascade,BrommerArxiv}. However, while many simulation parameters were set-up as to make them as comparable as possible, some details were overlooked. For instance, different randomly generated initial configurations  were used in each k-ART run, while the SEAKMC runs all started from the same initial state. In this study, all initial states are identical. Furthermore, the number of saddle-searches performed were not tracked in the original studies. As our results show, this is an important parameter for the 50-vacancy problem.

\subsubsection{Fe Intersitial Loops}

In previous work, SEAKMC was used to discover how two interstitial loops with $<111>$ and $<-1-11>$ Burgers vectors aggregate and transform to form either $<111>$,  $<-1-11>$ or $<100>$ loops \cite{xu2013solving}. Here, we perform these calculations using k-ART. We look at 37-intersitial loops in a 20-lattice parameter cubic box (for a total of 16074 atoms). The simulations are run at 573K, using the Ackland (2004) potential. In order to escape basins dominated by flickers and complete the transformation, we combined the bacMRM (with a 0.8 eV threshold to escape a super-basin) to a threshold on total RMSD displacement of each event (of 2.1 \AA~). Events with smaller displacements were ignored. This procedure ressembles that used to in Refs. \cite{jiang2014accelerated,scott2013sputtering} to study SiC and sputtering of Au surfaces with variations of k-ART. We start all simulations in a configuration where the two loops are interlocked (see Fig. \ref{fig:loop_pics}), which corresponds to a potential energy of -64404.05 eV.

\section{Results and Discussion}

\subsection{Comparison of saddle-search methods}
\subsubsection{Fe Mono-vacancy}\label{sec:compMonoVac}

\begin{table}[]
\caption{Benchmarks for 120 saddle-searches from stable Fe mono-vacancy configuration.}
\resizebox{0.5\textwidth}{!}{
\begin{tabular}{l|llll}
\hline
Search method & Dimer & Dimer & ARTn & ARTn \\
Initial displacement & Global & Local & Local & Local \\
Max steps in init. direction & 0 & 0 & 5 & 20 \\
Success rate (\%) & 100 & 99.2 & 100 & 99.2 \\
Execution time (s) & 37 & 29 & 30 & 50 \\
Saddles at 0.64 eV & 104 & 111 & 120 & 119 \\
Saddles at 1.07 eV & 14 & 6 & 0 & 0 \\
Other saddles & 2 & 2 & 0 & 
\end{tabular}}
\label{Tab:monovac1}
\end{table}

\begin{table}[]
\caption{Benchmarks for 120 saddle-searches from metastable Fe mono-vacancy configuration.}
\resizebox{0.5\textwidth}{!}{
\begin{tabular}{l|llll}
\hline
Search method & Dimer & Dimer & ARTn & ARTn \\
Initial displacement & Global & Local & Local & Local \\
Max steps in init. direction & 0 & 0 & 5 & 20 \\
Success rate (\%) & 98.3 & 99.2 & 89.2 & 90 \\
Execution time (s) & 126 & 73 & 78 & 111 \\
Saddles at 0.09 eV & 118 & 114 & 107 & 108 \\
Other saddles & 0 & 5 & 0 & 0
\end{tabular}}
\label{Tab:monovac2}
\end{table}

\begin{figure}
	\centering
    \includegraphics[width=8cm]{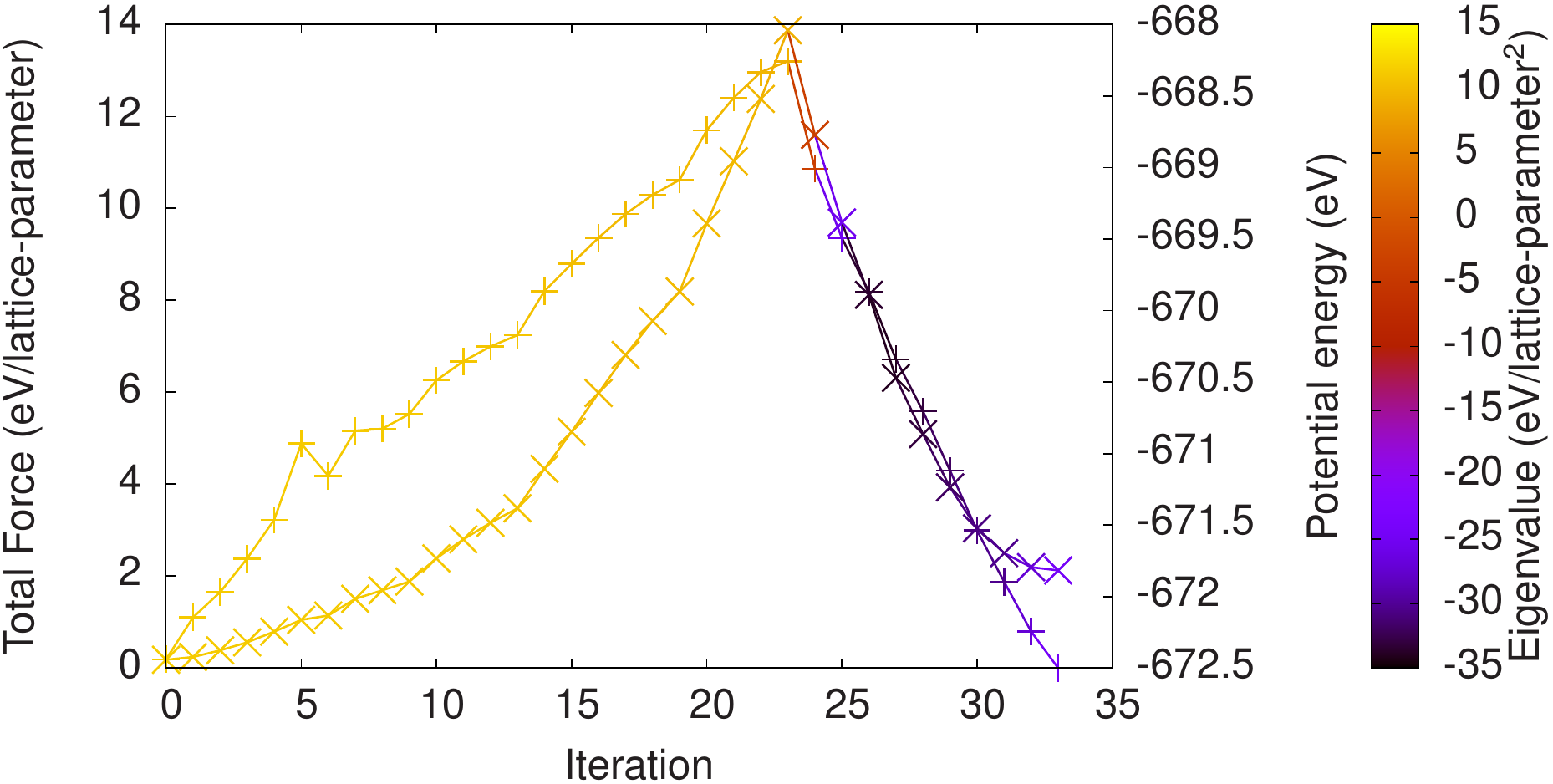}
    \includegraphics[width=8cm]{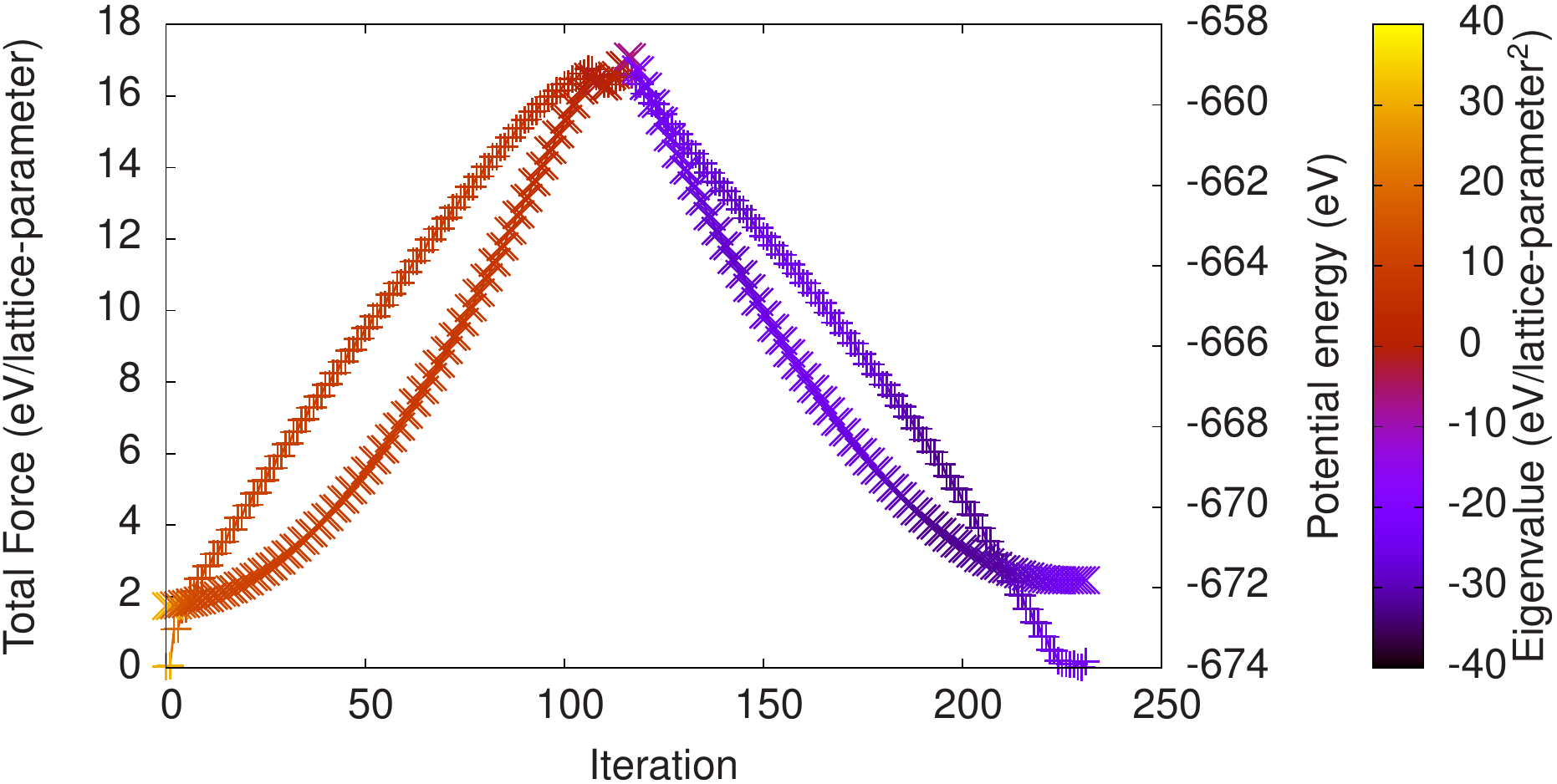}
     \includegraphics[width=8cm]{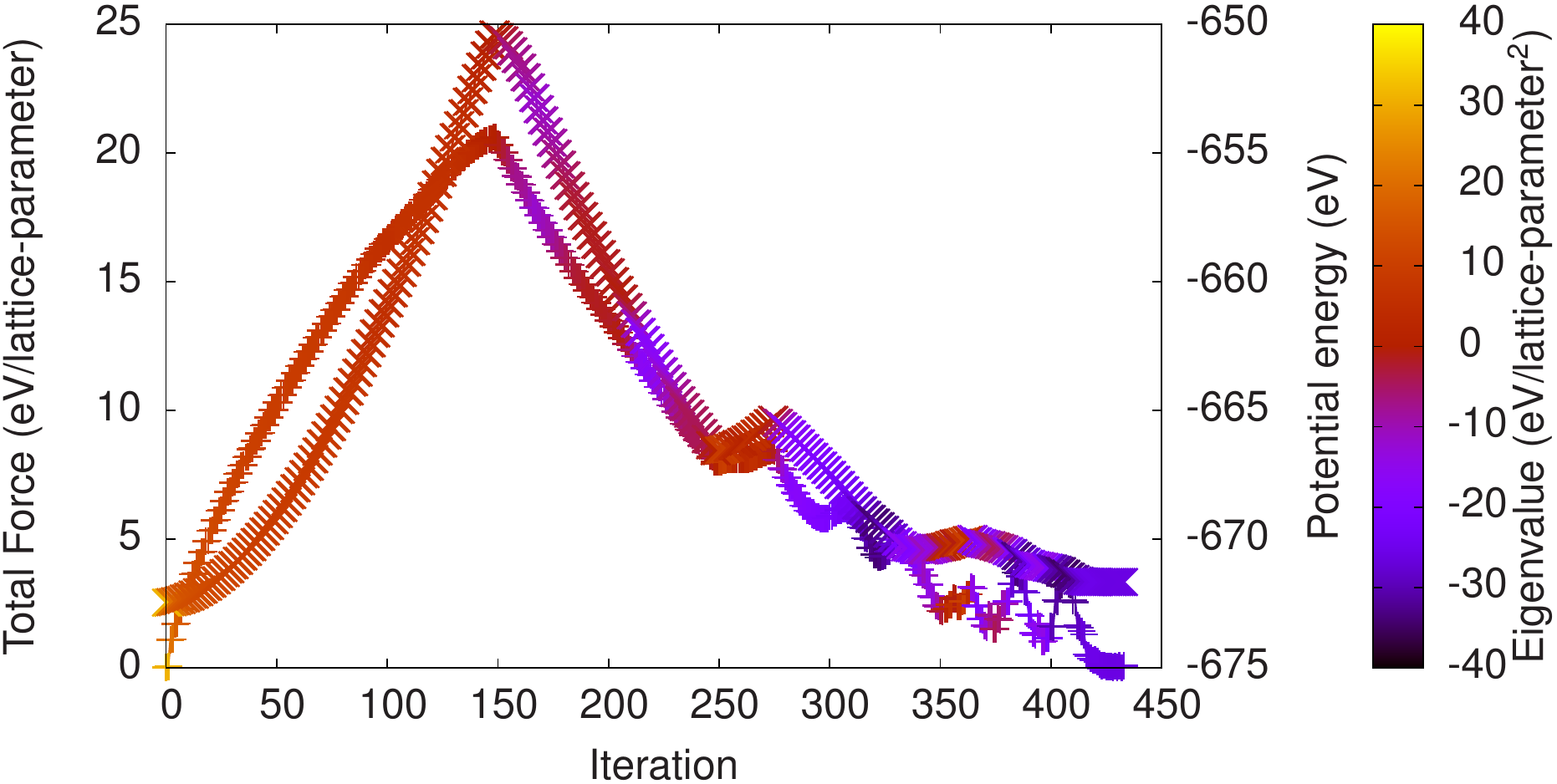}
    \caption{Total force (+) and potential energy (x) in typical saddle-searches for the Fe mono-vacancy system. The eigenvalue at each iteration is color-coded. The top figure corresponds to an ARTn search and the two bottom figures correspond to dimer searches. The the bottom figure differs because of a different random seed when initializing the saddle-search.} 
\label{fig:ARTn_vac_sad} 
\end{figure}

The results are shown in Tables \ref{Tab:monovac1} and \ref{Tab:monovac2}. We see that the methods both find the same activation barriers for diffusion, i.e. 0.64 eV from the stable state and 0.09 eV from the metastable state, in agreement with previous studies. The dimer method also finds a second saddle-point at 1.07 eV and other higher-energy saddle-points. This corresponds to an event where an atom moves to a position half-way between two empty lattice sites.  For both states, stable and metastable, the fastest method is the local dimer, closely followed by the ARTn implementation that limits the number of steps in the initial direction to 5. The global dimer search is slower than the local version. The ARTn searches with a maximum of 20 steps pushing in the initial direction is the slowest of the search methods.

We show examples of saddle-search profiles with ARTn and the dimer method in the Fe vacancy case in Fig. \ref{fig:ARTn_vac_sad}. These profiles are typical of the saddle-searches for all the systems reported in this paper. The dimer and ARTn profiles are very similar.  We see that the system's potential energy and forces increase until the eigenvalue of the softest mode becomes negative. Thereafter, both the potential energy and the forces decrease. This behavior does not seem to be dependent on the exact implementation of the saddle-search. Indeed, a FIRE+Lanczos-based implementation and a variant to the dimer method (see section \ref{Sub:method-saddle}) showed profiles very similar to those of ARTn and the standard conjugate-gradient-based dimer method. However, these two modified methods, that rely more heavily on relaxation in the perpendicular directions to the softest mode, often lead to searches that are trapped near the initial minimum, failing to converge to the saddle point.

One should note that a typical ART iteration is about 10 times more computationally expensive than a typical dimer iteration, because of the successive application of the Lanczos algorithm and tens of FIRE steps, compared to a few conjugate-gradient-based rotation iterations and translation attempts in a typical dimer-step. On the other hand, about 10 times less ARTn steps are needed to converge to the saddle.

We also notice that the time to converge to the saddle is related to the potential energy when a convexity is found. The higher the energy, the more iterations are necessary to reach the saddle-point. We also observe in the bottom panel of Fig. \ref{fig:ARTn_vac_sad} that convergence is more difficult if, during the minimization phase of the saddle-search, a concavity is hit. These correspond to the red-colored points around iteration number 175, number 250 and number 350. Nevertheless, by following the softest mode, the dimer method manages to find the saddle-point. However, one should keep in mind that hitting a concavity might indicate that we are no longer attracted to the saddle we were initially pursuing. This might lead to saddles that are not connected to the original minimum.

\subsubsection{Fe SIA}

\begin{table}[]
\caption{Benchmarks for 120 saddle-searches from Fe SIA configuration.}
\resizebox{0.5\textwidth}{!}{
\begin{tabular}{l|llll}
\hline
Search method & Dimer & Dimer & ARTn & ARTn \\
Initial displacement & Global & Local & Local & Local \\
Max steps in init. direction & 0 & 0 & 3 & 10 \\
Success rate (\%) & 87.5 & 98.3 & 98.3 & 98.3 \\
Execution time (s) & 401 & 54 & 48 & 51 \\
Saddles at 0.30-0.32 eV & 40 & 99 & 91 & 101 \\
Saddles at 0.41-0.42 eV & 5 & 16 & 27 & 16 \\
Saddles at 0.68 eV & 4 & 2 & 0 & 0\\
Other saddles & 56 & 1 & 0 & 1
\end{tabular}}
\label{Tab:SIA}
\end{table}

The results are shown in Table \ref{Tab:SIA}. We see that both methods find the 0.31 eV barrier for diffusion and the 0.41 eV barrier to rotate to a $1/2<111>$ configuration, in agreement with the literature. In this implementation, the dimer method samples more high-energy saddle-points. This is particularly true if the initial deformation is across the whole active volume. We have also tested this benchmark using a smaller step-size (70\% smaller, not shown in the table). While the execution was much slower (it was nearly four times slower), all searches resulted in 0.31 eV and 0.41 eV barriers.

Both implementations of ARTn perform better than the local dimer implementation (48 and 51 seconds vs 54 seconds, respectively). These three local deformation algorithms perform much better than the global deformation dimer, that is nearly 8 times slower and has a 11\% lower success rate than the local methods. 

The search profiles are similar to those of the mono-vacancy, plotted in Fig. \ref{fig:ARTn_vac_sad}. We see that searches that converge the slowest or that do not lead to a saddle are those that reach very high levels of potential energy and forces, which happens most often when globally deforming the active volume. We believe that the use of a bigger active volume for the SIA than the mono-vacancy (4 lattice-parameters vs 2.7) explains why a global deformation is so inefficient.

\subsubsection{50 Fe Vacancies}
\begin{table}[]
\caption{Benchmarks for 600 saddle-searches from Fe 50-vacancy configuration.}
\resizebox{0.5\textwidth}{!}{
\begin{tabular}{l|llll}
\hline
Search method & Dimer & Dimer & ARTn & ARTn \\
Initial displacement & Global & Local & Local & Local \\
Max steps in init. direction & 0 & 0 & 5 & 20 \\
Success rate (\%) & 100 & 100 & 100 & 99.5 \\
Execution time (s) & 166 & 126 & 156 & 260 \\
\end{tabular}}
\label{Tab:bench50}
\end{table}

\begin{figure}
	\centering
    \includegraphics[width=7cm]{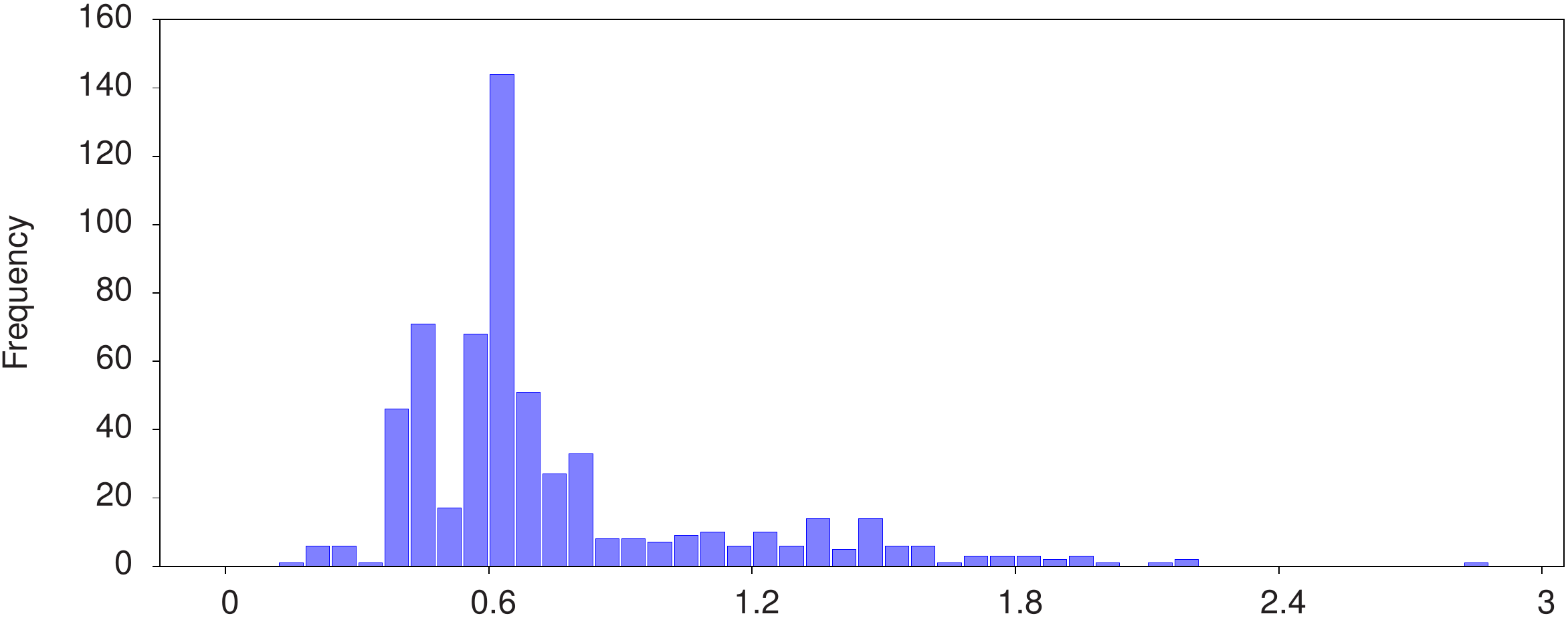} %global dimer
    \includegraphics[width=7cm]{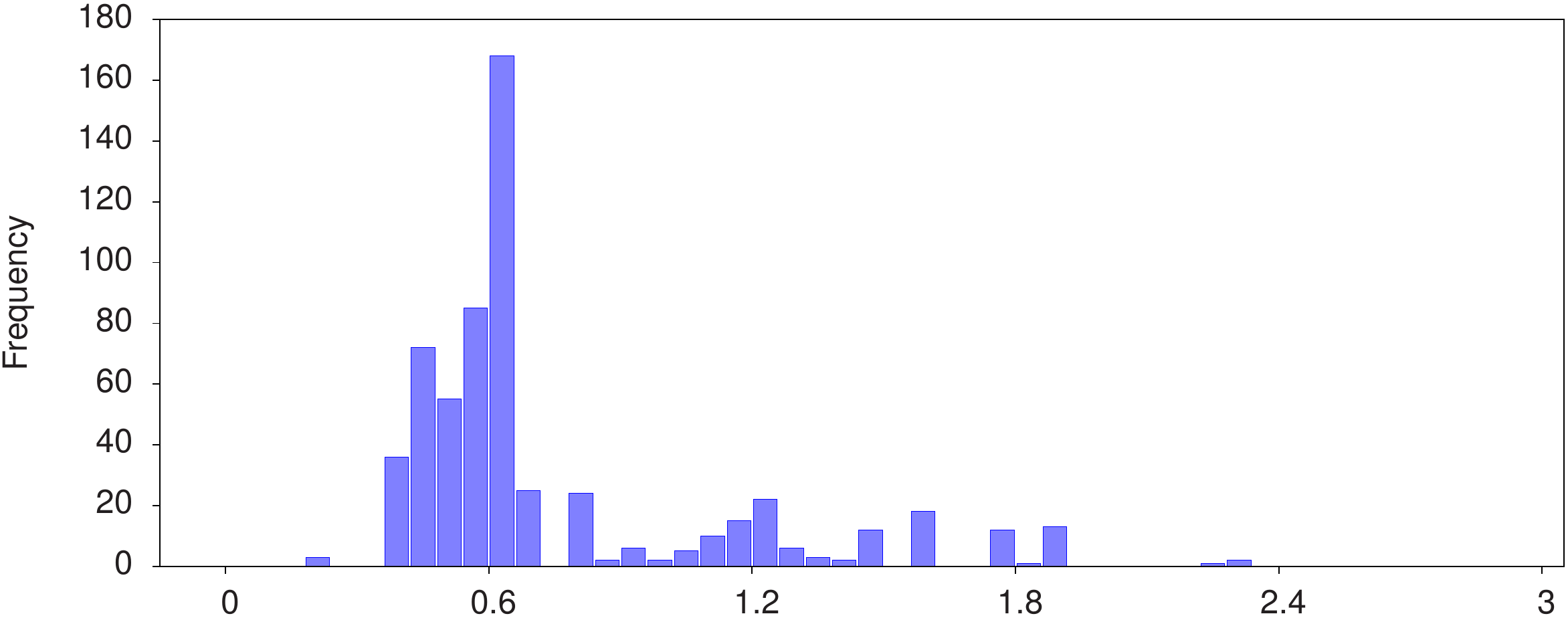} %local dimer
    \includegraphics[width=7cm]{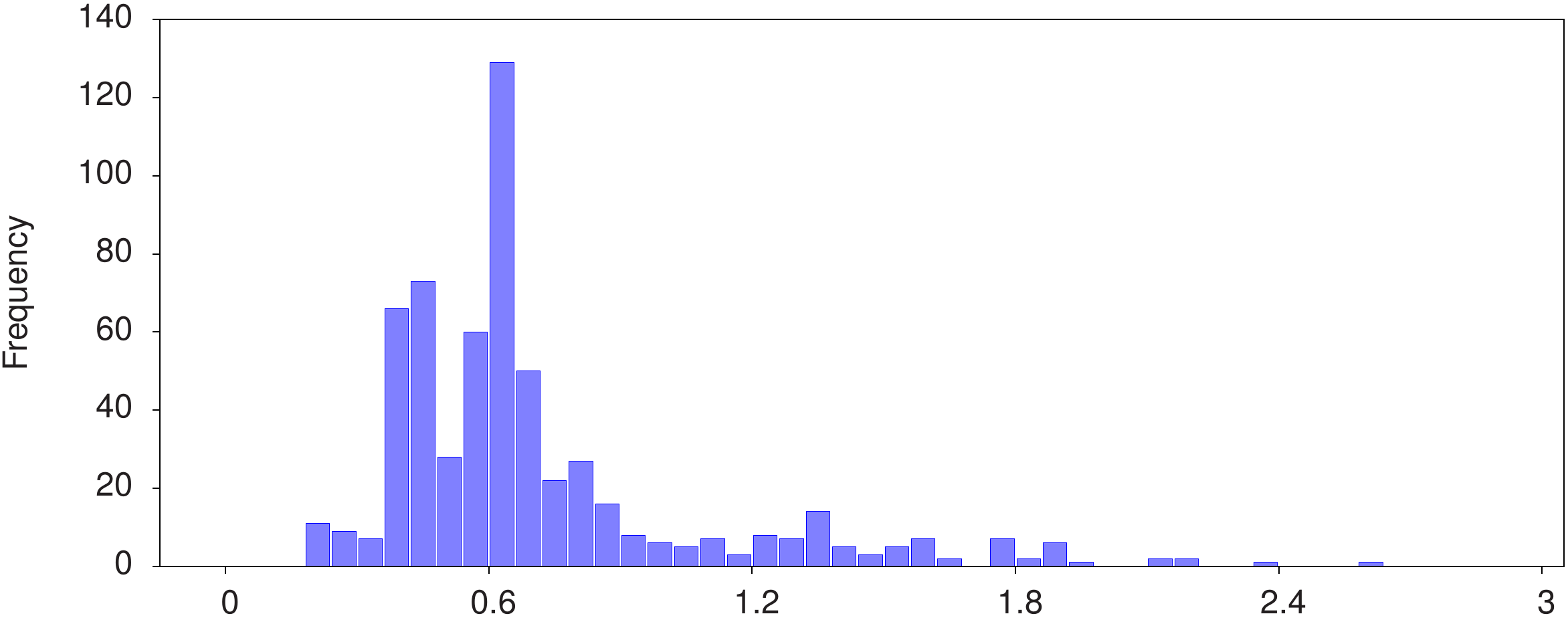} %5 kter ART
    \includegraphics[width=7cm]{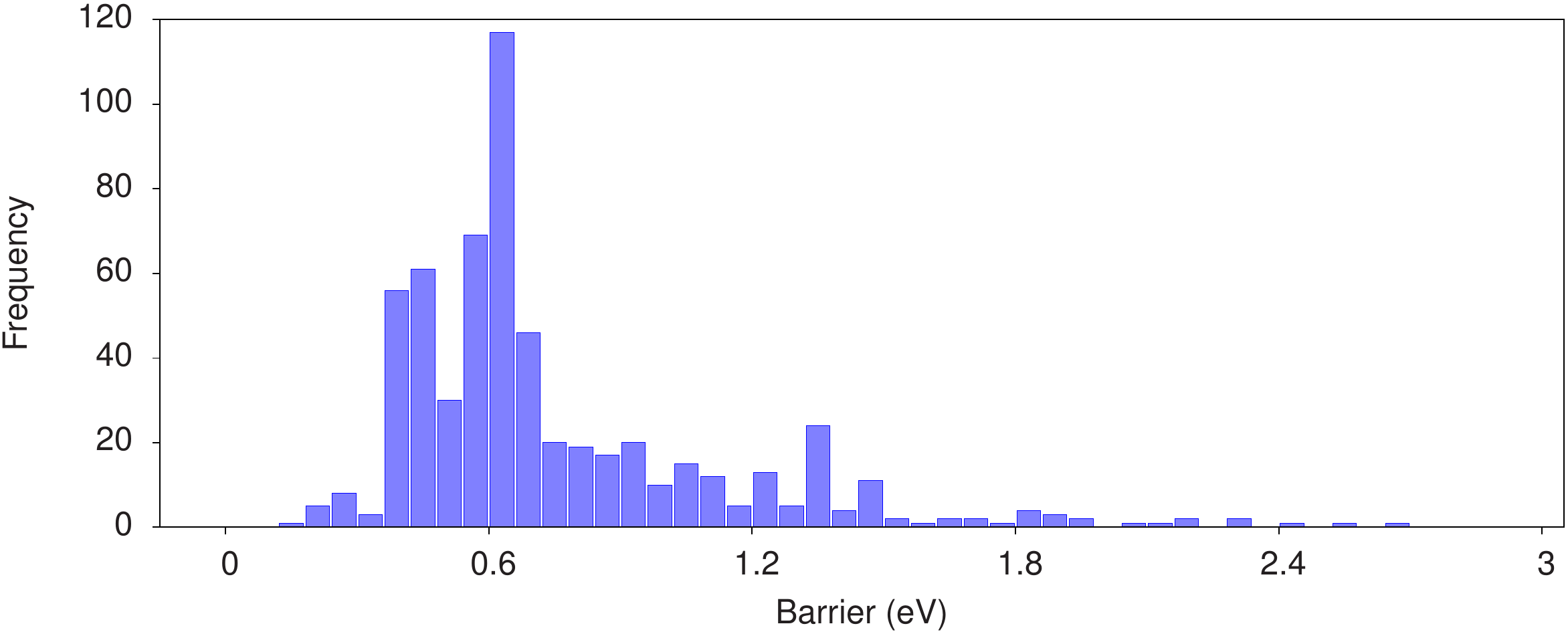}  %20ket ART
    \caption{Distribution of saddles found by the dimer method (top two panels) and ARTn (two bottom panels) for an Fe 50-vacancy configuration. From top to bottom: global dimer, local dimer, ARTn limited to 5 initial displacement steps, ARTn limited to 20 initial displacement steps.} 
\label{fig:50vac_histo} 
\end{figure}

The results are shown in Table \ref{Tab:bench50}. The fastest method is the local dimer (126 seconds), followed by the ARTn implementation that limits the number of steps in the initial direction to 5 (156 seconds). The global dimer search is slower than the local version (166 seconds). The ARTn searches with a maximum of 20 steps pushing in the initial direction is the slowest of the search methods (260 seconds). 

The distribution of saddles for each method is shown in Figure \ref{fig:50vac_histo}. They are very similar, no matter which method is selected. The local dimer method generates the most narrow distribution (with a peak frequency of 168) and the ARTn runs limited to 20 initial displacements steps have the broadest distribution (with a peak frequency of 117). Thus, while the local dimer generates saddle points more rapidly, its sampling is not as broad as that of other alternatives. As in the SIA case, the search profiles are similar to those of the mono-vacancy, plotted in Fig. \ref{fig:ARTn_vac_sad}.

We notice that the relative efficiency of the saddle-search implementations in this system is similar to the mono-vacancy system. Even though they interact, the energy landscape formed by the 50 vacancies has similarities with the landscape of a single vacancy.

\subsubsection{Fe interstitial loops}

\begin{table}[]
\caption{Benchmarks for 144 saddle-searches from an Fe configuration with two interlocked dislocation loops.}
\resizebox{0.5\textwidth}{!}{
\begin{tabular}{l|lll}
\hline
Search method & Dimer  & ARTn & ARTn \\
Initial displacement & Global  & Local & Local \\
Max steps in init. direction  & 0 & 3 & 20 \\
Success rate (\%) & 32.6  & 46.5 & 29.9 \\
Execution time (s) & 339  & 258 & 349 \\
\end{tabular}}
\label{Tab:benchloop}
\end{table}

\begin{figure}
	\centering
    \includegraphics[width=7cm]{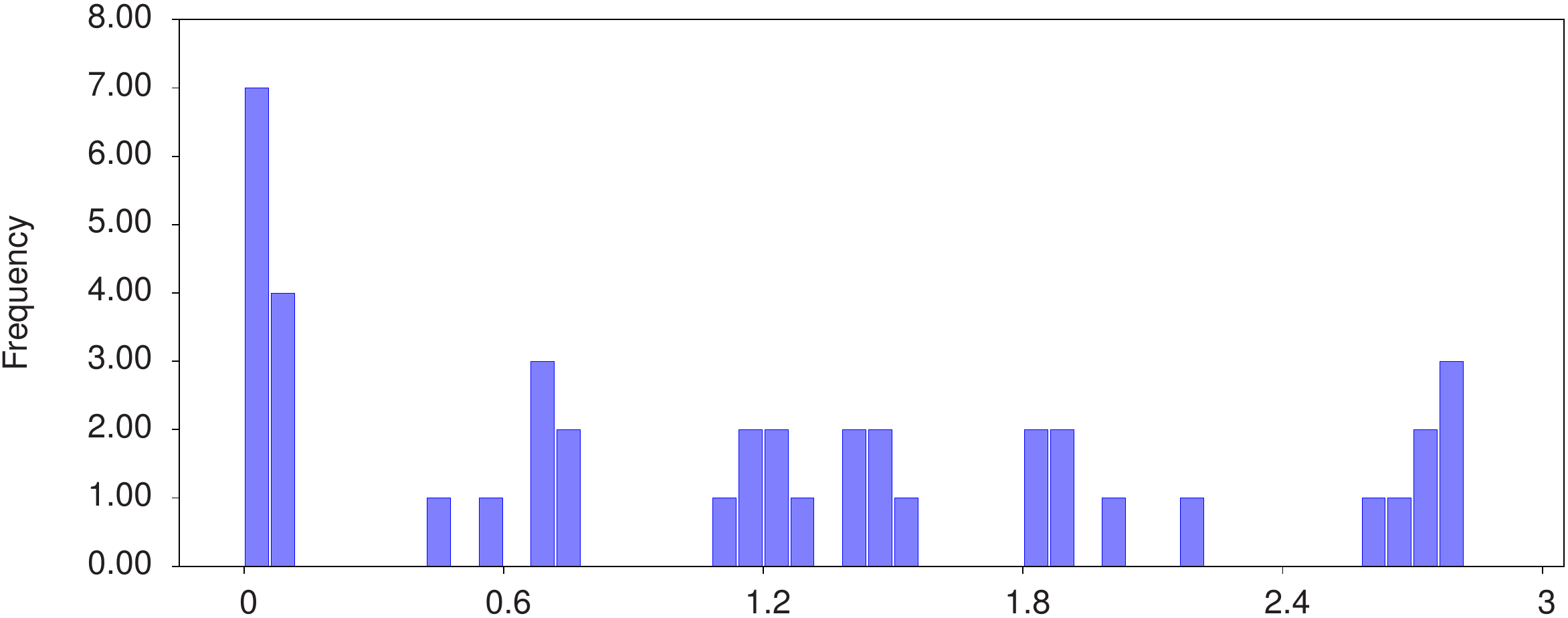} %global dimer
    \includegraphics[width=7cm]{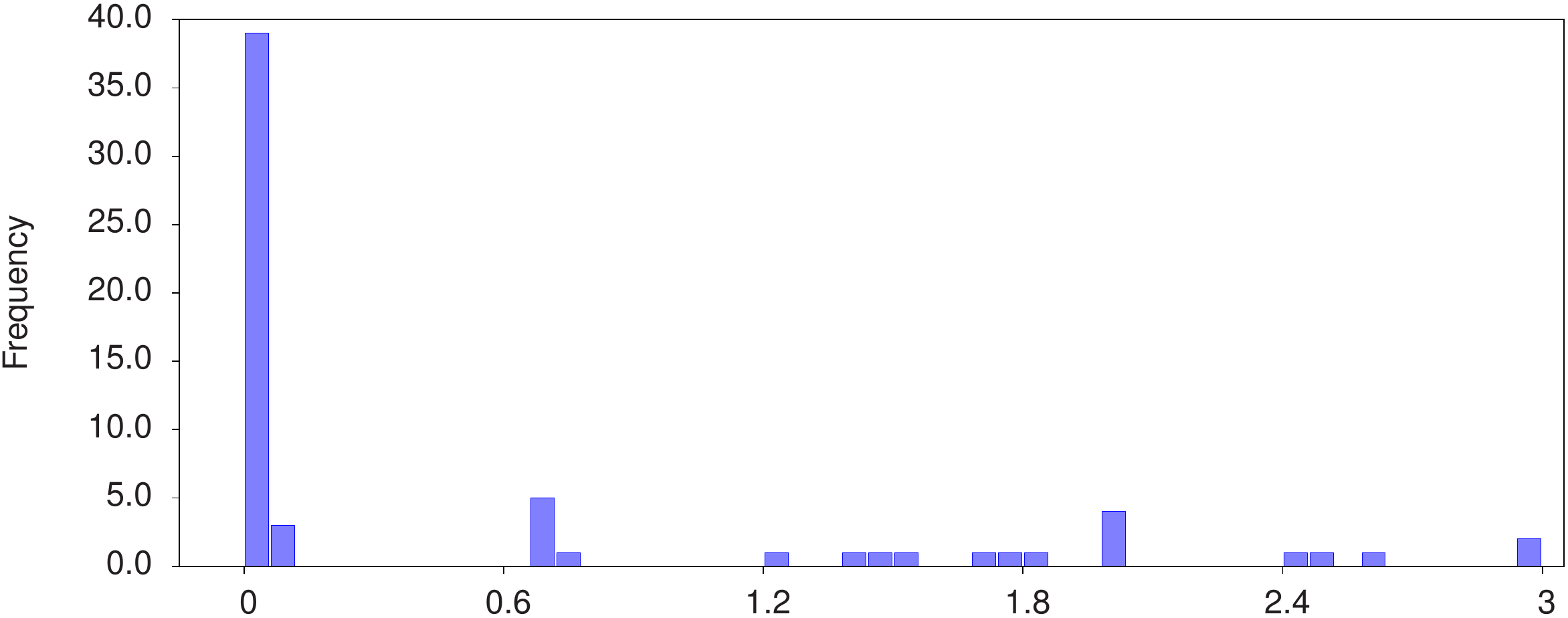} %local dimer
    \includegraphics[width=7cm]{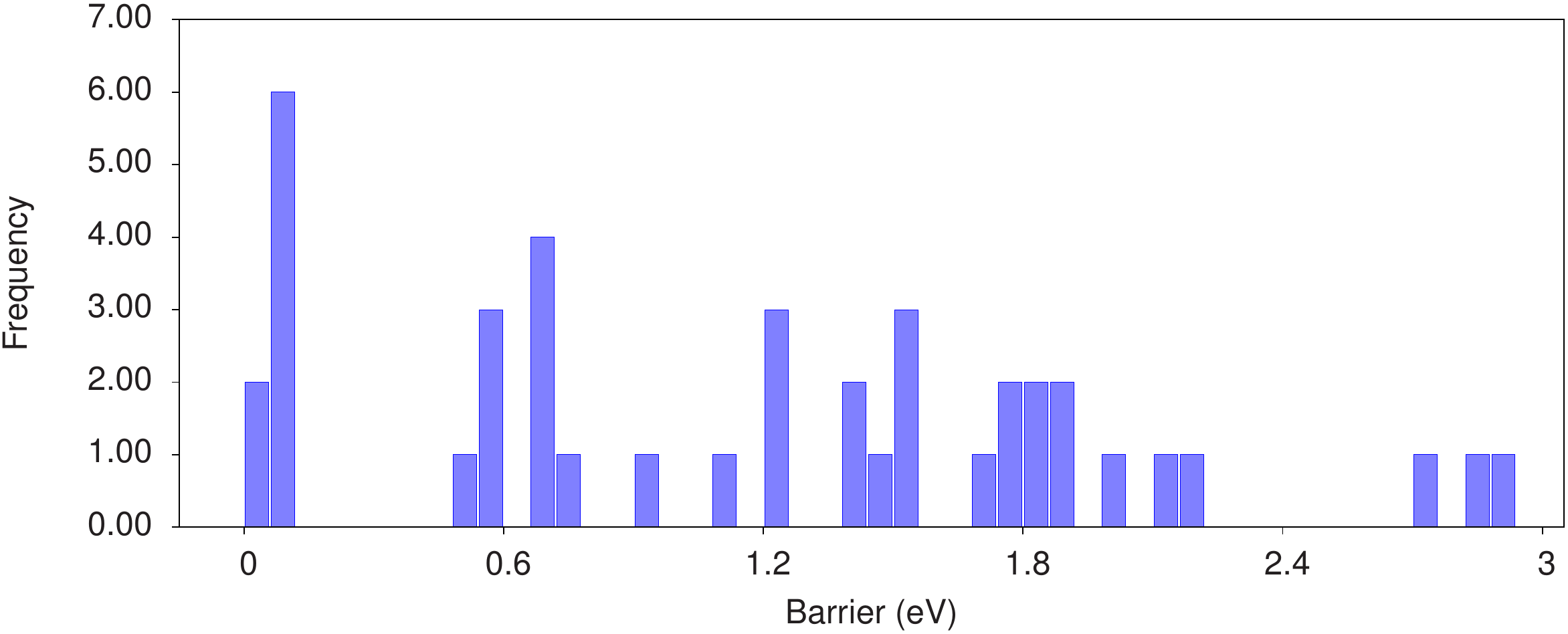} %5 kter ART
    \caption{Distribution of saddles found by the dimer method (top panel) and ARTn (two bottom panels) for the Fe interstitial loops system. From top to bottom: global dimer, ARTn limited to 3 initial displacement steps, ARTn limited to 20 initial displacement steps.} 
\label{fig:FeLoop_histo} 
\end{figure}

The results are shown in Table \ref{Tab:benchloop}. We did not implement a local version of the dimer method. The fastest method is the ARTn implementation that limits the number of steps in the initial direction to 3 (258 seconds). The global version of the dimer does better than the ARTn implementation that limits the number of steps in the initial direction to 20 (339 seconds vs 349 seconds).

In Fig. \ref{fig:FeLoop_histo}, we show the distribution of saddles found by the three methods. Overall, the global dimer version and the ARTn runs that limit initial displacements steps to 20 are very similar. They each show a double peak at low barriers with 2-7 events (between 0 and 0.06 eV) and many peaks of size 1 to 4 saddles at higher energies. When limiting ARTn searches to 3 steps in the initial direction, we observe a much larger peak at low-energy (between 0 and 0.03 eV), with 39 events, and a few occurrences at high energies. This latter implementation seems to favor low-energy barriers.

In this case, localizing the initial deformation does not seem to greatly affect the efficiency of the search. One should note that both the ARTn and dimer searches lead to states where the whole volume has displaced atoms at the saddle-point, which may explain why they perform similarly.

\subsubsection{FeCr interstitial loops}
\begin{table}[]
\caption{Benchmarks for 144 saddle-searches from an FeCr configuration with two interlocked dislocation loops.}
\resizebox{0.5\textwidth}{!}{
\begin{tabular}{l|lll}
\hline
Search method & Dimer  & ARTn & ARTn \\
Initial displacement & Global  & Local & Local \\
Max steps in init. direction  & 0 & 3 & 20 \\
Success rate (\%) & 5.6  & 32.6 & 29.9 \\
Execution time (s) & 2248  & 383 & 381 \\
\end{tabular}}
\label{Tab:benchloopFeCr}
\end{table}

\begin{figure}
	\centering
    \includegraphics[width=7cm]{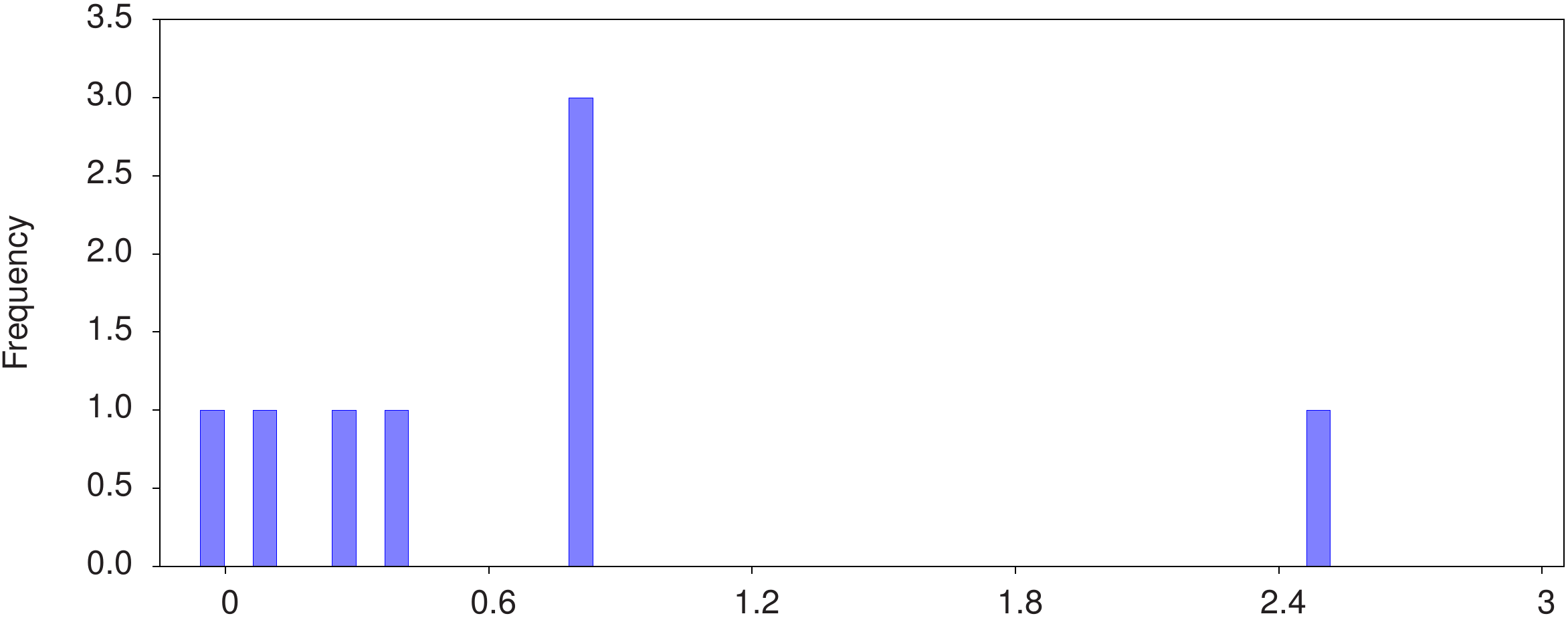} %global dimer
    \includegraphics[width=7cm]{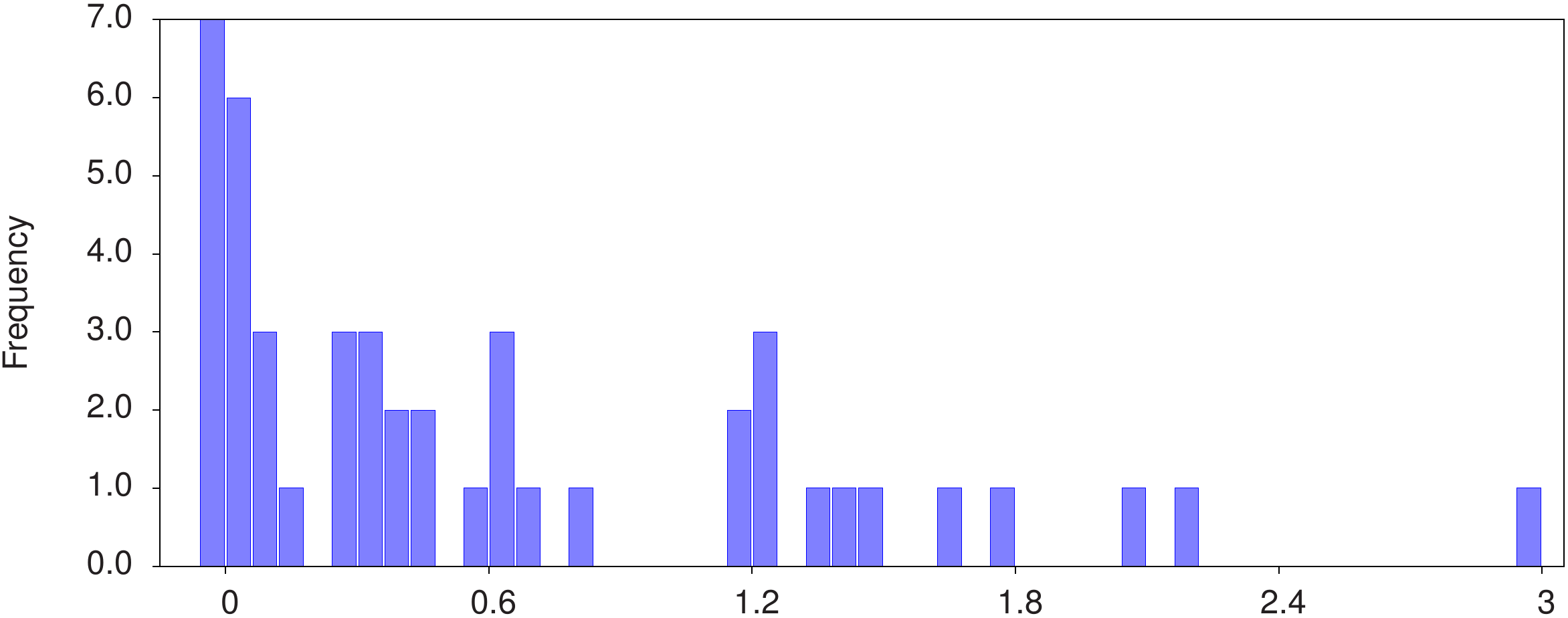} %local dimer
    \includegraphics[width=7cm]{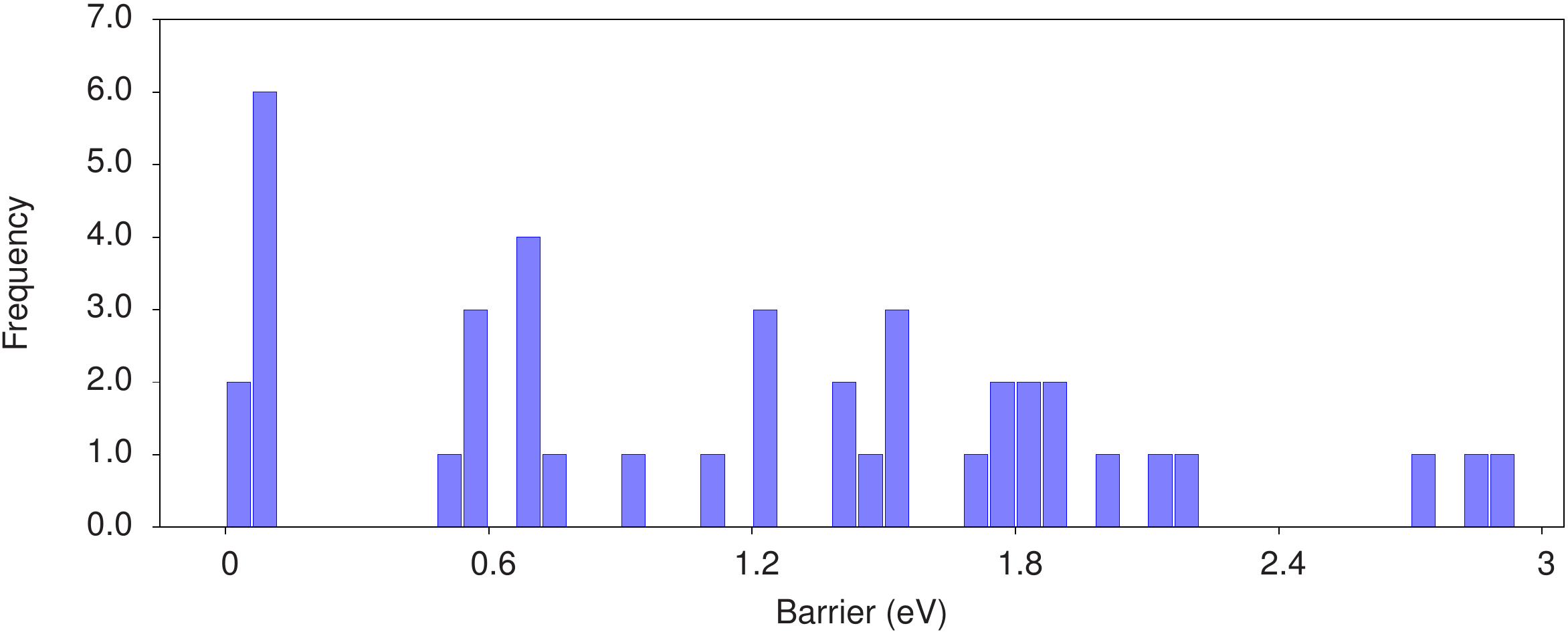} %5 kter ART
    \caption{Distribution of saddles found by the dimer method (top panel) and ARTn (two bottom panels) for the FeCr interstitial loops system. From top to bottom: global dimer, ARTn limited to 3 initial displacement steps, ARTn limited to 20 initial displacement steps.} 
\label{fig:FeCrLoop_histo} 
\end{figure}

The results are shown in Table \ref{Tab:benchloopFeCr}. We did not implement a local version of the dimer method. The fastest method is the ARTn implementation that limits the number of steps in the initial direction to 20 (381 seconds), closely followed by the ARTn implementation that limits the number of steps in the initial direction to 3 (383 seconds) . The global version of the dimer does much worse than ARTn (2248 seconds). The dimer also has a much smaller success rate (5.6\%) than ARTn (about 30\%).

In Fig. \ref{fig:FeCrLoop_histo}, we show the distribution of saddles found by three methods. The dimer method only found 8 events, scattered across the whole spectrum. The two ARTn implementation have somewhat similar spectra. When limiting ARTn searches to 3 steps in the initial direction, we observe more low-energy events (from 0 to 0.6 eV). This latter implementation seems to favor low-energy barriers, as in the pure Fe case.

The addition of Cr to the system seems to create problems when activating the whole volume. This might be related to the presence of optical phonons, which are absente in the pure Fe, and to the many-body nature of the two-band model, which can add roughness to the second derivatives of the potential energy surface. The effect of these vibrational modes and sharp changes in forces on the eigenmode-following search are probably enhanced when deforming the lattice rapidly over a large radius.

\subsection{On-the-fly KMC comparison}

\subsubsection{Fe 50-vacancy problem}

\begin{figure}
	\centering
    \includegraphics[width=9cm]{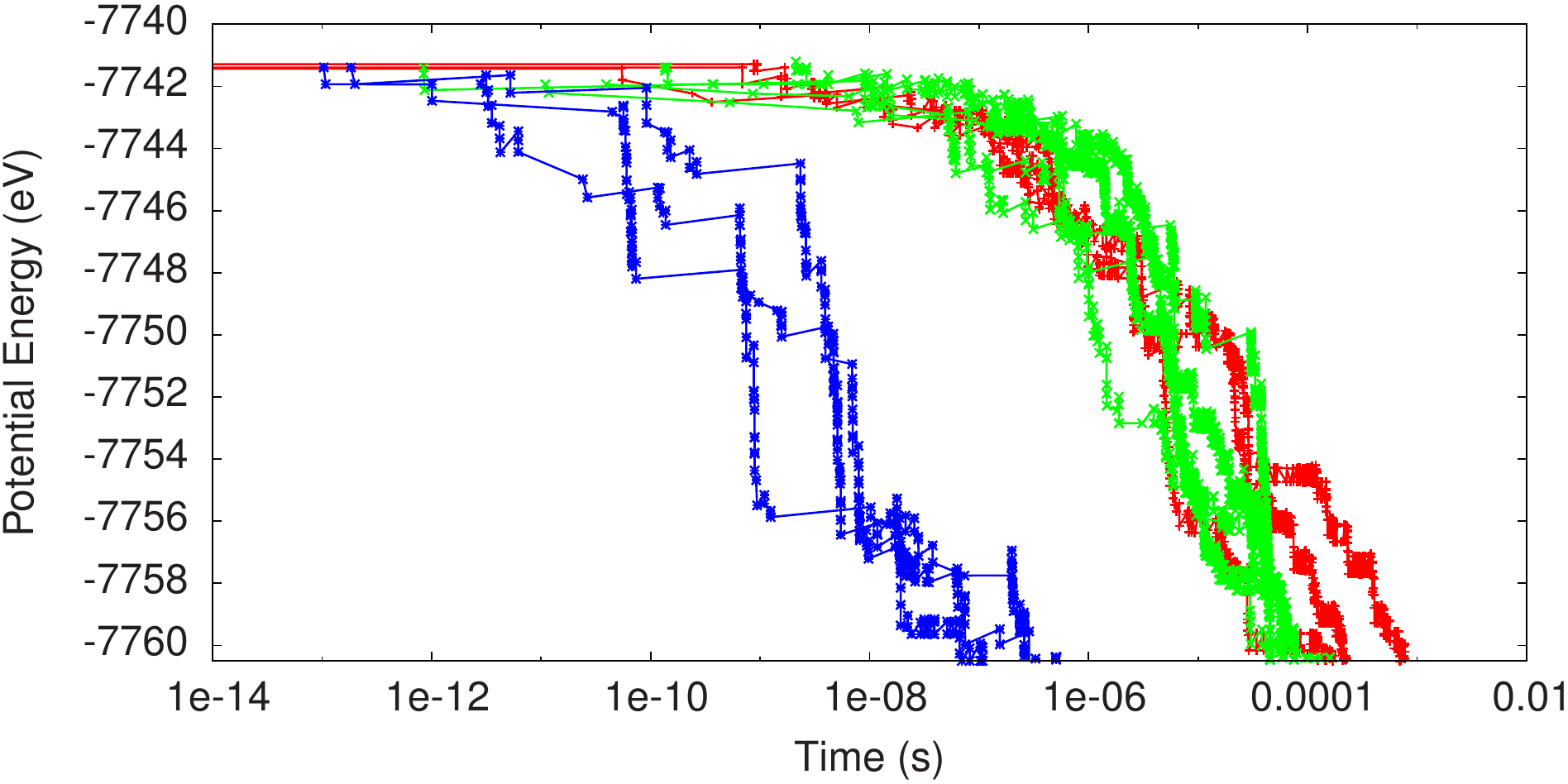} %global dimer
    \caption{Potential energy of 50 mono-vacancies aggregating at 323K. Eight of these curves have a large overlap.The three red curves correspond to k-ART runs and the five green curves correspond to SEAKMC runs that sample 2 events per defect per step. The three blue curves on the left that do not overlap with the others are SEAKMC runs that sample 24 events per defect per step} 
\label{fig:50Vruns} 
\end{figure}

We tracked the generation of saddle points in k-ART and SEAKMC when simulating the aggregation of 50 vacancies in a 1950-atom Fe box. A notable feature of SEAKMC is that it has an overall success rate of saddle-convergence of 95\%. This is possible because searches are centered directly on the crystallographic defects. In comparison, k-ART has a success rate of 35\% for generic-event generation (before analysis to exclude multiple identical events) and a 90\% specific-event generation success rate.This is because the saddle-searches on not centered on crystallographic defects, but rather on atoms with a connectivity graph different from that of an atom in a perfect crystal. As a consequence, while there are 50 defects(vacancies) in the system, the NAUTY-based topology analysis identifies, on average, 250 different topologies in the system, which must all be explored by k-ART. Simple geometrical considerations imply that most topologies represent atoms that sit far from the defects, near the limit of the region where atoms are included in the topology-analysis, i.e. a sphere with a radius of 5.6 \AA. Centering searches on these atoms is not likely to generate events, especially if the displacements are not limited to an Active Volume. As for failures of specific-event generation, this is related to failures to execute the correct geometrical transformation from the generic event's coordinates to a specific atomic site, which relies on the graph canonical representation, which can be problematic for configurations where inversion symmetries play a role.

In such an off-equilibrium system, the kinetics are largely determined by the presence of high-energy barriers that separate the system from relaxation events (i.e. the replenish-and-relax model). If one misses a low-lying energy barrier that can eventually lead to relaxation and instead chooses to execute a high-energy barrier, the kinetics will be unphysically slowed down. Good sampling of the potential energy landscape between relaxation steps (i.e. the aggregation of a vacancy with a cluster) is crucial for good kinetics.

Considering that 20 generic searches per new topology were launched in k-ART, that the total number of topologies visited during the runs is about 8600, that the number of KMC steps for each run is roughly 700 and a 35\% success rate, we estimate that, on average, approximately 85 events were generated every step. Taking into consideration the new events created by the refinement of specific events, this is comparable to the 100 event searches (2 searches per defect) launched in SEAKMC. 

In Fig. \ref{fig:50Vruns} , we show five SEAKMC runs with 2 searches per defect and three k-ART runs, that are stopped after the system reaches a potential energy of -7760 eV, which roughly corresponds to the energy when half the mono-vacancy remain. All the simulations are in good agreement. We note that some SEAKMC runs permitted the execution of events with negative energy barriers. Although these are unphysical within TST, they usually correspond to instabilities in the potential energy. The good agreement of the runs in Fig. \ref{fig:50Vruns} indicate that executing these events does not play an important role in determining the kinetics.

Furthermore, we tried two different activation volumes (2.7 \AA~ and 4.0\AA). This does not seem to affect the aggregation time. However, we noticed that the two runs with an activation volume of 2.7 \AA~ took about 1600 steps to reach -7760 eV, while those with a larger active volume took about 300 steps. This is probably due to the fact that we used a global dimer search. Indeed, when initial deformations are global, the saddle search involves configurations of higher energy, as we explain in Sec. \ref{sec:compMonoVac}. The larger the active volume, the greater the effect. This favors finding saddles with higher energies, which, if using a limited sampling of events, can help escape flickers. Indeed, we tried such a simulation using the ARTn, which performs localized initial deformations, within SEAKMC (not shown). We limited the number of steps in the initial direction to 5, as to favor low-energy barriers as much as possible, and found that the system oscillates between states separated by low-barriers for a long time.

It thus appears that using an active volume and a global dimer method indirectly permits handling flickers by favoring events with larger barriers and displacements. Although, unlike the bacMRM, this is not a rigorous method, the results in Fig. \ref{fig:50Vruns} seem to indicate that it produces the right kinetics. It also indicates that the two different sets of approximations implemented in k-ART and SEAKMC, described in Table \ref{Tab:Approximations}, produce very similar kinetics, at least for this system. We note that the use of a too aggressive basin threshold leads to slower relaxation (not shown in the figures), because it leads to less sampling and thus can miss low-barrier relaxation events. 

From the standpoint of computational efficiency, the SEAKMC took roughly 70 cpu-hours to reach -7760 eV, while the k-ART runs took 2500. This is mostly due to the use of the Active Volume, which permits force calculations across a much smaller number of atoms (165 vs 1950). The rest of the acceleration is associated with the larger success rates of saddle-searches. We note that for a system of such size, implementing localized forces calculations that dynamically include new participants, as described in Refs. \cite{beland2011kinetic,joly2012optimization}, does not produce a significant speedup, because of the computational overhead of such a tactic.

We also ran a group of three simulations that sample 1200 events per step (24 events per defect), twelve times more than the other SEAKMC and k-ART simulations. Because of the large cost of such sampling, we only performed them with SEAKMC. The results show that the kinetics are 1000 times faster. Even if we assume that adding 12 times more events to the catalog (thus increasing the rates by a similar proportion) amounts to oversampling, the kinetics are nearly 100 times faster at this level of sampling. These faster runs correspond to the aggregation time observed in Ref. \cite{xu2013cascade}. This reconciles the apparent mismatch between previous publications.

\begin{figure}
	\centering
    \includegraphics[width=9cm]{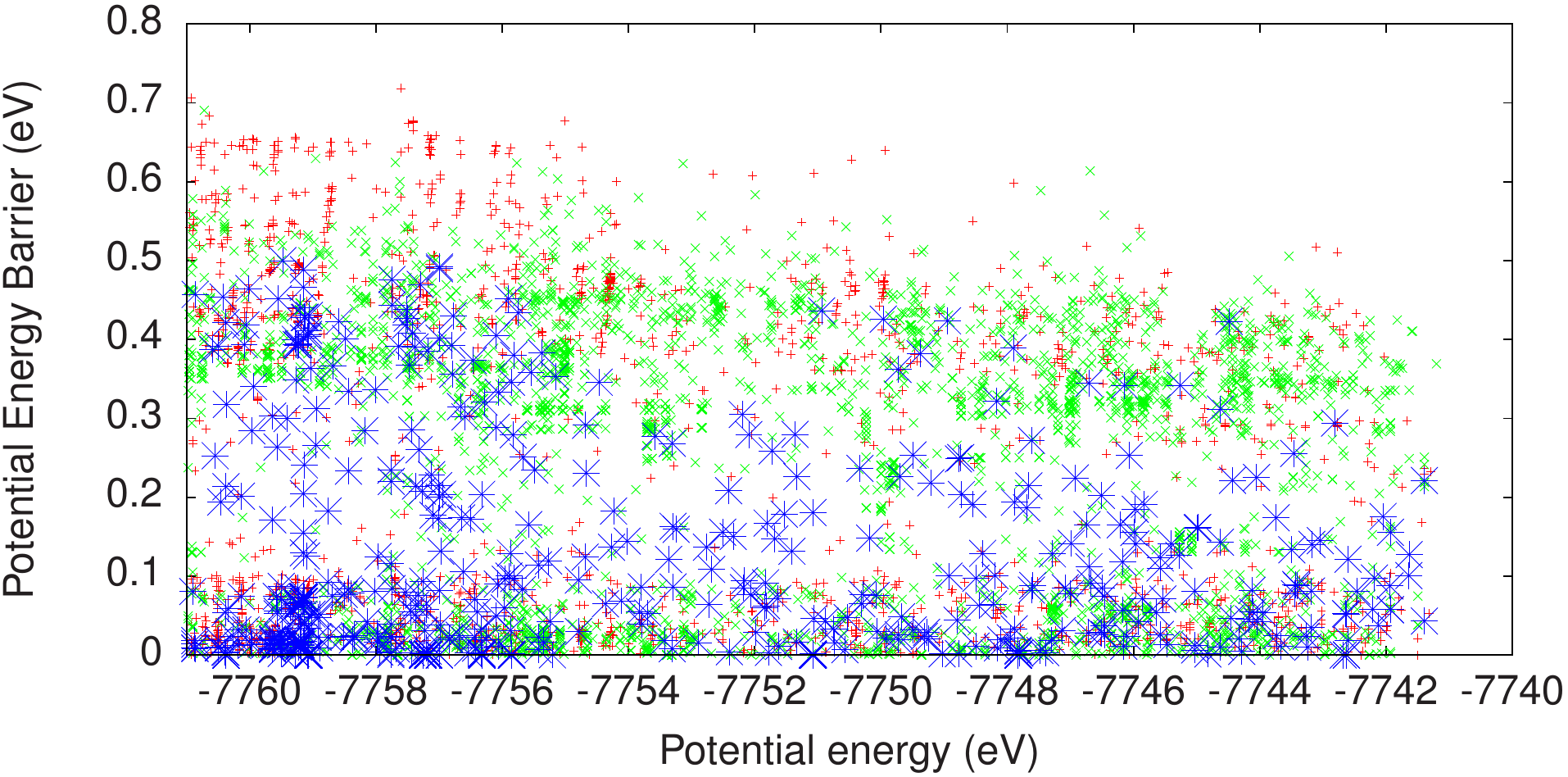} %global dimer
    \caption{Potential barriers energy of executed events of 50 mono-vacancies aggregating at 323K as a function of the potential energy of the local minimum. The red (+) points are k-ART runs, the green (X) points are SEAKMC runs sampling 2 events per defect per step and large blue stars are SEAKMC runs sampling 24 events per defect per step.} 
\label{fig:50Vbarriers} 
\end{figure}

Fig. \ref{fig:50Vbarriers} explains this behavior. It shows the barriers that are executed as a function of the potential energy of the local minimum. Since the potential energy decreases nearly-monotonously with time, reading this graph from right to left gives a measure of the activation barriers that characterize relaxation. We see that the k-ART and SEAKMC runs with comparable sampling (100 events per step) execute events with similar barriers. The SEAKMC runs that sample 1200 events per step are characterized by smaller barriers (roughly 0.25 eV lower) for all configurations. In other words, by sampling more, the algorithm found events with lower barriers that nonetheless permit large displacements without many flickers. This is an indication that we were in fact under-sampling the system in the other runs.

\subsubsection{Fe Interstital-loop problem}

\begin{figure}
	\centering
    \includegraphics[width=3.3cm]{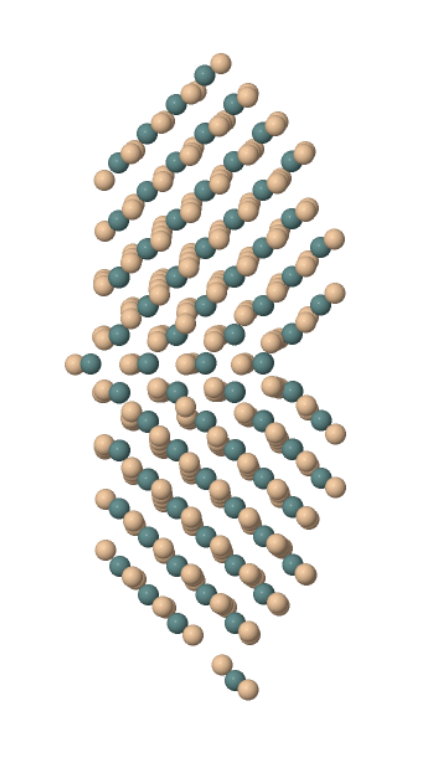} %global dimer
    \includegraphics[width=4.2cm]{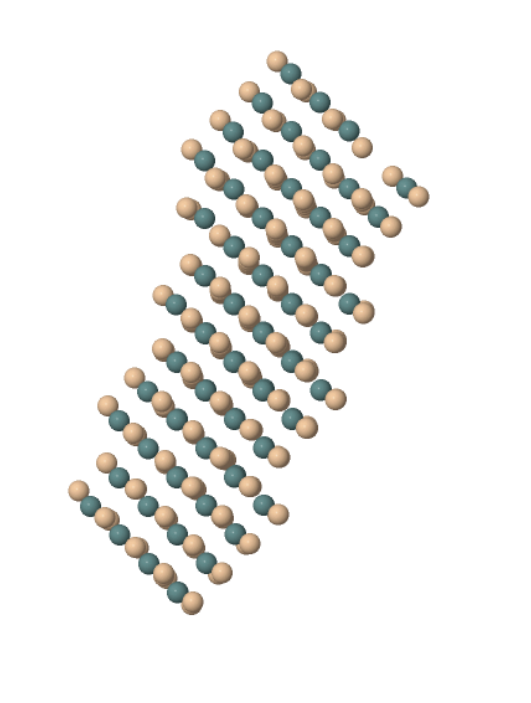}
    \includegraphics[width=3cm]{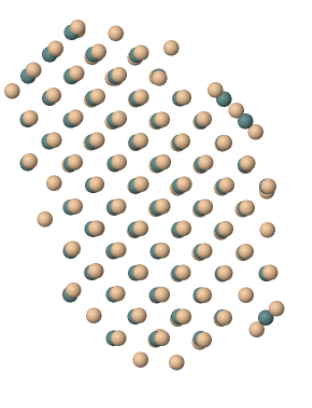}
    \caption{Illustration of the initial state for the Fe interstitial-loop runs (upper left) and two outcomes: transformation to a $<111>$ loop (upper right) and to a $<100>$ loop (bottom). All illustrations are from the $<100>$ perspective. Beige spheres are interstitials and blue spheres are vacancies.} 
\label{fig:loop_pics} 
\end{figure}

We show illustrations of the initial configuration for our runs (two $1/2<111>$ loops) and of two outcomes after a transformation (a $1/2<111>$ loop and a $<100>$ loop) in Fig. \ref{fig:loop_pics}. In order to complete the transformation in a shorter amount of states, as explained in the methodology, all events with an RMSD smaller than 2.1 \AA~ are ignored. If events with lower displacements, e.g. 1.0 \AA, are included in k-ART, flickering motions are created, leading to super-basins with several hundred states before a transformation can be executed. By imposing the threshold on displacements, transformations were executed in k-ART within 140 to 190 KMC steps, not including flickers treated by the bacMRM. We note that this is larger than the number of steps in the SEAKMC runs described in Ref. \cite{xu2013solving}, that took less than 100 KMC steps to execute. We believe that this is related to the use of global initial deformations, which favors events with larger barriers and displacements. We also note that sampling a large number of events per step (more than 120) also leads to flickering in SEAKMC. In other words, SEAKMC handles the flicker problem by imposing large initial deformations in the saddle search and by limiting the sampling of the system at every step.

\begin{figure}
	\centering
    \includegraphics[width=8cm]{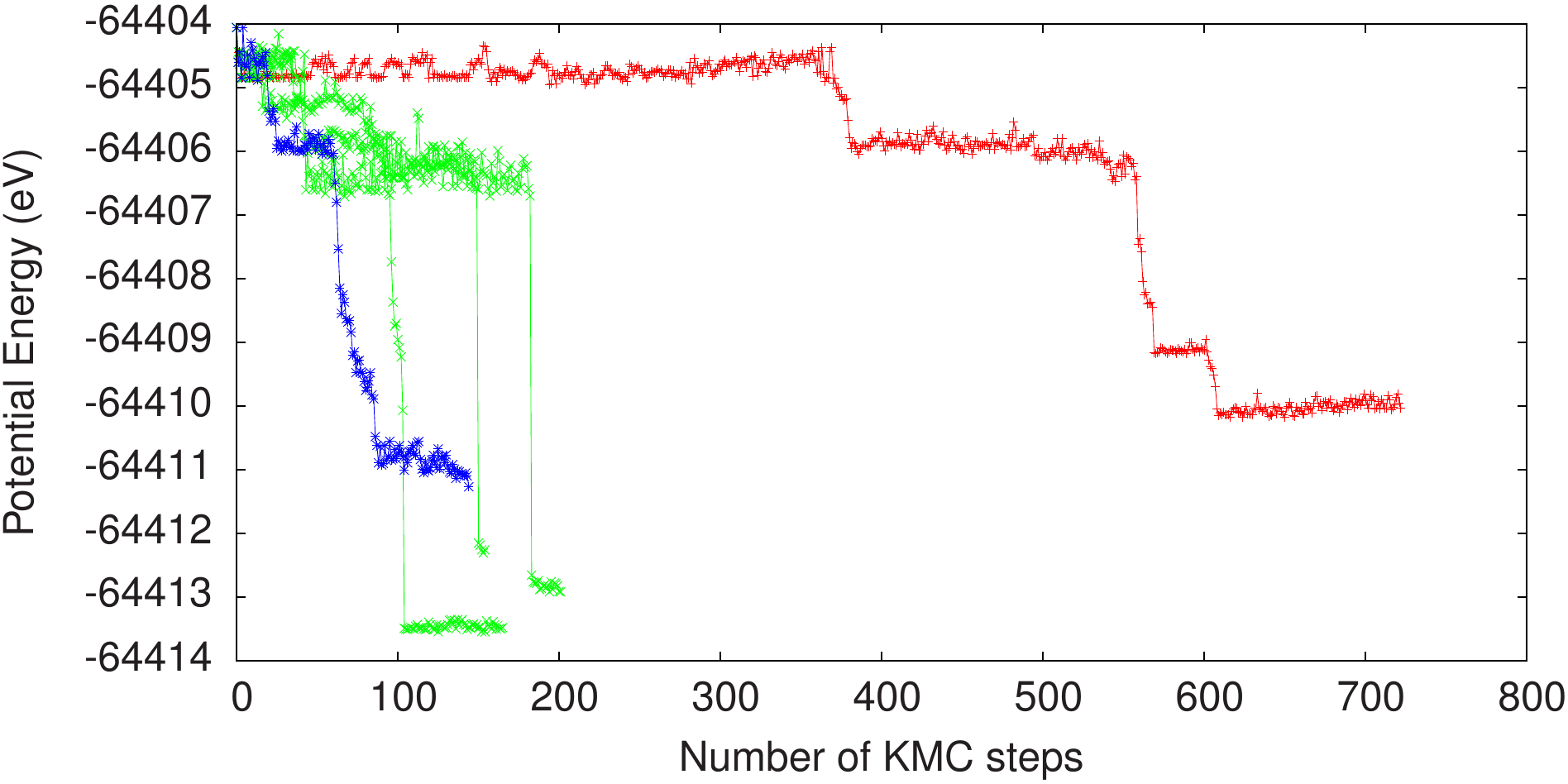} %global dimer
    \caption{The potential energy of Fe loop system as a function of the number of KMC steps. The red line (+) corresponds to a k-ART run with a RMSD displacement threshold of 1.0\AA~ on events (see text) that leads to a $<100>$ loop. The other runs have a 2.0\AA~ displacement threshold. The green (X) lines lead to a $1/2<111>$ loop, while the blue (stars) leads to a $<100>$ loop. } 
\label{fig:loop_time} 
\end{figure}

Both k-ART and SEAKMC predict transformations to $<100>$ and $1/2<111>$. We show the potential-energy evolution during these runs in Fig.\ref{fig:loop_time}. Over these five runs, the highest barrier crossed was 0.81 $\pm$ 0.23 eV.  In Ref. \cite{xu2013solving}, averaging  over more than one hundred SEAKMC simulations, the authors found 0.73 $\pm$ 0.11 eV. Furthermore, the atomistic mechanism that these two algorithms describe were confirmed through MD simulations. Overall both methods come to similar conclusions, even though they each use a very different set of approximations. 

The k-ART simulations took 2000-2500 cpu-hours to execute the transformation, while SEAKMC took about 71 cpu-hours. This ratio is in fact similar to that found for the 50-vacancy case. This is mostly due to the use of the Active Volume, that permits force calculations to be limited to 1050 atoms, compared to 16074 atoms in the case of k-ART. We note that we were able to implement a local force calculation algorithm for this potential within k-ART (as described in Refs. \cite{beland2011kinetic,joly2012optimization}), but the overall speed was comparable to the use of the optimized force computation library of IMD, as implemented in Ref.\cite{beland2011kinetic}.

\section{General Remarks}

\subsection{Saddle-Search Methods}
In order to converge quickly to the saddle, one should attempt to limit the potential energy and forces of the system when it reaches the convexity of the potential energy surface. A straightforward way to do this is to limit the number of atoms which are involved in the initial deformation of the system. Indeed, in our benchmarks, localized initial deformations have led to faster convergence to the saddle-point than deformations across the whole active volume. Also, limiting the number of steps in the initial random displacement with ARTn can accelerate convergence in some systems, such as the mono-vacancy, the 50-vacancy and the pure Fe interstitial-loop systems presented here. However, this has not helped increased convergence in the SIA case, nor the FeCr interstitial-loop case.

This general assessment, that locally constraining initial displacements generally leads to faster saddle-searches and good sampling of exit states, was noted and quantified in previous studies \cite{pedersen2011efficient,pedersen2014bowl}, albeit it was not shown to apply both to the dimer method and ARTn, nor specifically to Fe and FeCr defects.

The increase and subsequent relaxation of the potential energy indicate that these methods do not follow the minimum energy path (MEP) to the saddle-point. One might think that this is related to an insufficient relaxation in the directions perpendicular to the softest eigenmode. However, the same behavior was observed when modifying the algorithms to emphasize perpendicular relaxation. Why is this? The literature provides some insights to this effect. Take, for instance, Refs. \cite{henkelman1999dimer,mauro2005simplified,samanta2012atomistic}. The first is the original dimer paper and the other two describe the eigenvector-following method and the Gentlest-Ascent Dynamics (GAD), which are closely related to the dimer and ARTn methods. These studies present tests on simple, 2-dimensional potentials. In all case, if the initial deformation is not sufficiently parallel to the MEP, the potential energy will follow an increase and subsequent decrease when searching for a saddle-point. In many-dimensions, is it virtually impossible to generate a random displacement that has a large parallelity with the MEP, thus explaining why we observe profiles similar to those illustrated in Fig. \ref{fig:ARTn_vac_sad}.

We note that Ref. \cite{samanta2012atomistic} also provides an alternative, open-ended, method to find saddle-points that closely follow the MEP, GAD molecular dynamics (GAD-MD). In this case, one generates an approximation of the eigendirection of lowest curvature using the variational principle and runs a molecular dynamics simulation, applying $\mathbf{F}=\mathbf{F}_T-2\mathbf{F}_p$, as in the dimer method and the GAD. At every time step, the approximation of the eigendirection is updated using the Euler method. This special bias of the potential destabilizes the energy landscape's minima and stabilizes the saddle-points, while favoring low-lying free-energy-states. In their test on a 2-dimensional surface, the trajectories seem to stay close to the MEP. We also note that in Ref. \cite{henkelman1999dimer}, the authors suggest that dimer searches should start from MD configurations in order to make the trajectories as physically relevant as possible. In that same spirit, we believe that executing GAD-MD until a convexity is discovered and then starting an eigenmode-following method might permit convergence to saddle points relatively efficiently while staying near the MEP. However, such a method may favor low-energy saddles even more than do traditional methods, which may make sampling less efficient. While such an implementation is out of the scope of this work, such development would be of interest, at least from a pedagogical perspective, even if not from the perspective of kinetics simulation.

\subsection{On-the-fly KMC}

It seems that both sets of approximations implemented in k-ART and SEAKMC lead to similar kinetics. We showed that SEAKMC treatment of flickers implicitly depends on the use of the global dimer method and somewhat under-sampling the system. While this has worked well for the two systems described in this study, it is quite possible it might not work in other cases. For example, our calculations in FeCr indicate that using the global dimer method is not adequate. Preliminary simulations of interstitial loops in FeCr show that using such a technique widely overestimates the activation barriers.

It is preferable, in our opinion, to use saddle-search methods based on local deformations (or MD) and afterwards discard events which will probably not be important to the kinetics (such as events with very small symmetric barriers and events with small RMSD displacements). This should increase the chance of finding low-barrier events that lead to significant changes in the system. This type of procedure seems widespread in the literature. Another approach is to massively increase the temperature in the hope that higher-energy barriers will be selected and exit the flickering super-basin \cite{jiang2014accelerated}.

However, discarding events in order to minimize flickering is a procedure that can significantly distort kinetics. Two methods exist in kinetic-ART to solve this problem: TABU and the bacMRM. In Ref. \cite{GowonouArxiv}, we show that these methods generate very similar kinetics. If possible, one should use the bacMRM, since it is a more rigorous approach. If super-basins are very large, this might not be the case, and the use of TABU may be preferable to discarding low-displacement events.

We also note that many of the algorithms are not fundamentally tied to all the approximations described in Table \ref{Tab:Approximations}. For example, detailed balance and checks for duplicity of events can be implemented in SEAKMC through bookkeeping. Similarly, the performance gap between k-ART and SEAKMC could be narrowed by introducing the active volume approximation in k-ART. We should note that in Si, k-ART's local force implementation led to essentially no cost increase with system-size \cite{joly2012optimization} above 5000-atoms. In such systems, the active volume advantage would probably be much smaller.

Another strong feature of k-ART is its proven ability to simulate the kinetics of amorphous materials. It is not clear if the Active Volume approximation is appropriate for these highly disordered materials. In addition, while the topology-based criteria that coordinates saddle-searches and cataloging is completely independent of crystallinity, \emph{ad hoc} procedures for coordinating saddle-searches would probably need to be implemented in SEAKMC. Further work is needed to clarify this open issue.

\section{Conclusion}

In this work, we have have compared and contrasted two state-of-the-art aKMC methods, k-ART and SEAKMC. Firstly, we looked at the relative performance of the dimer method and ARTn, two open-ended saddle-search methods in simulations related to point defects, interacting vacancies and interstitial loops in Fe and FeCr. Secondly, we compared the kinetics of aggregation of 50 vacancies in a 1950-atom Fe box. Finally, we compared the kinetics of transformation of $1/2<111>$ loops in Fe to either a $<111>$ loop or a $<100>$ loop.

Generally, the performance of the dimer method and ARTn are similar if one ensures that the initial displacements during a search are limited to a localized subset of atoms. This generally permits faster convergence, a better success rate, and increases the probability of finding low-barrier saddle-points, which are of crucial importance for kinetics.

k-ART and SEAKMC simulations of the 50-vacancy system are in agreement if one is careful to generate the same number of events per KMC step. We also discovered that under-sampling by a factor twelve leads to aggregation 1000 times slower, through the discovery of higher-barrier events that lead to relaxation. The two algorithms also draw the same picture for the kinetics of the dislocation loops. 

This confirms that the different set of approximations that these algorithms implement were appropriate for this class of systems. It also indicates that both methods could be improved by adapting some elements from one another. For example, k-ART could significantly accelerate simulations by introducing active volumes, while the use of the bacMRM in SEAKMC would make it more rigorous when the energy landscape is dominated by flickers.

\section{acknowledgements}

LKB thanks Alexander V. Barashev for insightful discussions.

Research at the Oak Ridge National Laboratory sponsored by the U.S.
Department of Energy, Office of Basic Energy Sciences, Materials
Sciences and Engineering Division, "Center for Defect Physics," an
Energy Frontier Research Center. LKB acknowledges a fellowship awarded
by the Fonds Qu\'eb\'ecois de recherche Nature et Technologies.

k-ART is available upon request to Normand Mousseau and SEAKMC is available upon request to the authors.

\bibliography{Comparebib}

\end{document}